\documentclass[tighten,twocolumn,times]{aastex63}

\usepackage{epsfig,amsmath,amsfonts,amssymb,graphicx,subfigure,enumitem,url,natbib}

\received{........}
\revised{........}
\accepted{........}
\shorttitle{Linear polarization toward the cluster Alessi 1}
\shortauthors{Singh et al.}

\begin{document}

\title{Optical Linear Polarization toward the Open Star Cluster Casado Alessi 1}

\email{ssingh@aries.res.in} 
\email{jeewan@aries.res.in}

\author{Sadhana Singh}
\affiliation{Aryabhatta Research Institute of Observational Sciences (ARIES), Manora Peak, Nainital 263001, India}
\author{Jeewan C. Pandey}
\affiliation{Aryabhatta Research Institute of Observational Sciences (ARIES), Manora Peak, Nainital 263001, India}
\author{R. K. S. Yadav}
\affiliation{Aryabhatta Research Institute of Observational Sciences (ARIES), Manora Peak, Nainital 263001, India}
\author{Biman J. Medhi}
\affiliation{Gauhati University, Guwahati 781014, India}
\begin{abstract}
We present \textit{B-}, \textit{V-}, \textit{R-}, and \textit{I-}bands linear polarimetric observations of 73 stars in the direction of open star cluster Casado Alessi 1 (hereafter Alessi 1). We aim to use polarimetry as a tool to investigate the properties and distribution of dust grains toward the direction of the cluster.  The polarimetric observations were carried out using the ARIES IMaging POLarimeter mounted at the 104 cm telescope of ARIES, Nainital (India).  Using the \textit{Gaia} photometric data the age and distance of the cluster are estimated to be $0.8\pm0.1$ Gyr and $673\pm98$ pc, respectively.  A total of 66 stars with a 26\arcmin\, radius from the cluster are identified as members of the cluster using the astrometric approach. Out of these 66 members, 15 stars were observed polarimetrically and found to have the same value of polarization. The majority of the stars in the region follow the general law of the polarization for the interstellar medium, indicating that polarization toward the cluster Alessi 1  is dominated by foreground dust grains. The average values of the maximum polarization ($P_{max}$) and the wavelength corresponding to the maximum polarization ($\lambda_{max}$) toward the cluster are found to be $0.83\pm0.03$ \% and $0.59\pm0.04$ $\mu$m, respectively. Also, dust grains toward the cluster appear to be aligned, possibly due to the galactic magnetic field.

\end{abstract}


\section{Introduction}\label{sec:intro}
The presence of dust grains in the interstellar medium (ISM) can be revealed by interaction with starlight and then further emission/scattering from the dust grains, which in turn produce polarized light. The dichroic extinction of starlight by aligned asymmetric dust grains in the ISM is considered the main source of ISM polarization. Although the identity of the dominant grain alignment mechanism has proved to be an intriguing problem in grain dynamics \citep{1997ApJ...490..273L,2003JQSRT..79..881L,2012JQSRT.113.2334V}, it is generally believed that asymmetric grains tend to become aligned to the local magnetic field through the grains shortest axis parallel to the magnetic field \citep{1951ApJ...114..206D}. Thus, linear polarimetric observations may provide information about the geometry of the underlying magnetic field of the region.  This will shed light on the observed region's type, e.g., whether it is   star-forming, a dense molecular cloud, etc. \citep[see][]{1996ApJ...462..316H,2007JQSRT.106..225L,2018arXiv180602806F}. The wavelength dependence of polarization gives information about the size distribution of dust grains. The total to selective extinction is thought to be related to the wavelength corresponding to maximum polarization ($\lambda_{max}$) \citep{1978A&A....66...57W}. Thus, the polarimetric technique is considered a useful tool to explore the properties of dust grains \citep[e.g.][]{1994ApJ...431..783K,2011EAS....52..259V}. Polarimetric study of open star clusters is useful for determining the properties of foreground interstellar dust, as the majority of the clusters have basic information like distance, membership, color, etc.  \citep[e.g., see ][]{2008MNRAS.391..447F,2008MNRAS.388..105M}. 

 \begin{table*}
	\centering
	\caption{Polarization and Position Angles of  Polarized Standard Star HD19820 on Two Different Nights}
	\label{tab:standard}
	\begin{tabular}{lcccccc} 
		\hline		
	\multicolumn{5}{c}{Present Work} & \multicolumn{2}{c}{\citet{1992AJ....104.1563S}}\\	
	
	\multicolumn{3}{r}{2017 November 22}&
	\multicolumn{2}{c}{2017 November 23}\\
	\hline
	Passband & $P(\%)$ & $\theta\left({^o}\right) $& $P(\%)$ & $\theta\left({^o}\right) $ & $P(\%)$ & $\theta\left({^o}\right) $\\
  	\hline
	\textit{B} & 4.71$\pm$0.02 & 115.8$\pm$0.1 & 4.65$\pm$0.10 & 115.6$\pm$0.6  & 4.699$\pm$0.036 & 115.7$\pm$0.2\\
	\textit{V} & 4.76$\pm$0.11 & 114.9$\pm$0.6 & 4.43$\pm$0.15 & 114.8$\pm$0.9 & 4.787$\pm$0.028 & 114.93$\pm$0.17\\
	\textit{R} & 4.45$\pm$0.11 & 114.6$\pm$0.7 & 4.43$\pm$0.13 & 114.5$\pm$0.8 & 4.526$\pm$0.025 & 114.46$\pm$0.16\\
	\textit{I} & 3.94$\pm$0.07 & 114.4$\pm$0.5 & 4.13$\pm$0.07 & 114.1$\pm$0.5 & 4.081$\pm$0.024 & 114.48$\pm$0.17\\
	\hline
				
	\end{tabular}
\end{table*}
In this paper, we have studied an open star cluster Alessi 1 [R.A.(J2000) = $00^{h}53^{m}27^{s}$, Decl.(J2000) = $+49^{o}34\arcmin11\arcsec$; $l$ = $123^{o}.26$, $b$ = $-13^{o}.30$] \citep{2005A&A...438.1163K} using the \textit{B-}, \textit{V-}, \textit{R-}, and \textit{I-}band polarimetric observations and \textit{Gaia} archival data. According to \citet{2003A&A...410..565A}, Alessi 1  has a large diameter of 54\arcmin\, with an isolated group of stars.  They found 28 probable members of the cluster from their kinematic analysis. Using photometric data from the Tycho 2 catalog, they did not find any cluster sequence in the color-magnitude diagram (CMD). \citet{2004AN....325..740K} estimated the distance to be 800 pc, the angular radius of core and cluster to be 7\arcmin.8 and 27\arcmin\,, the age to be 0.7 Gyr and 23 most probable members of the cluster. However, \citet{2013A&A...558A..53K} estimated the distance to the cluster to be nearly 750 pc, and found the age of the cluster to be similar to that found by \citet{2004AN....325..740K}. They also estimated the angular radii  of the core, central part, and the cluster for  Alessi 1 as 3\arcmin\,, 10\arcmin.5, and 25\arcmin.5, respectively. \citet{2019A&A...623A.108B} have estimated age, distance modulus, extinction in the \textit{V} and \textit{G} bands as 0.86 Gyr, 9.255 mag, 0.322 mag, and 0.314 mag, respectively, for the cluster Alessi 1. Recently, \cite{2018A&A...618A..93C} derived the membership probability of 63 stars in this cluster using \textit{Gaia} data, and found 56 stars have membership probabilities greater than 50\%.

The paper is organized as follows. The observations and data reduction, along with the archival \textit{Gaia} data, are described in  Section \ref{sec:obs}, and in Section \ref{sec:analy}, we present our results, analysis and discussion. In Section \ref{sec:sum}, we summarize our results.

\section{Observations and data reduction} \label{sec:obs}
\subsection{Polarimetry}
The optical polarimetric observations of the cluster Alessi 1 were carried out on 2017 November 22 and 23 in the  \textit{B}, \textit{V}, \textit{R}, and \textit{I} photometric passbands using the ARIES IMaging POLarimeter  \citep[AIMPOL;][]{2004BASI...32..159R} mounted at the Cassegrain focus of the 104 cm Sampurnanand telescope of ARIES, Nainital. The telescope is a Ritchey-Chretien reflector with a focal ratio of \textit{f/}13 \citep{1972oams.conf...20S}. The detector was a Tek 1k $\times$ 1k charge-coupled device (CCD) camera cooled by liquid nitrogen.  Each pixel of the CCD corresponds to $1\arcsec.73\times1\arcsec.73$ resulting in an entire field of view (fov) of CCD of $\sim 8$\arcmin\, in diameter. The gain and read-out-noise of the CCD are 11.98 e$^-$ per ADU and 7.0 e$^-$, respectively. Observations were taken in four positions of the half-wave plate (HWP), $\alpha$, of  $0^{o}$, $22^{o}.5$, $45^{o}$, and $67^{o}.5$ from north-south to get the linear polarization. There was a 27 pixel separation between the ordinary and extraordinary images of each source in the CCD frame. As the fov of the AIMPOL is small, therefore to observe a large fov for the cluster,  we have divided the cluster into seven regions. Thus,  we observed a total fov of $\sim 12$\arcmin\, in radius, which is similar to the radius of the central part of the cluster as estimated by \citet{2013A&A...558A..53K}.  Three continuous observations were taken in each position of HWP for every region. The exposure times were varied from  50 to 180 s depending upon the filter used. In order to improve the signal-to-noise ratio,  we have added all three frames in each position of HWP. Stars HD 19820 and HD 21447 were observed as standard polarized and unpolarized stars, respectively, in the \textit{B}, \textit{V}, \textit{R}, and \textit{I} filters for the calibration on each night of observation.

The Image Reduction and Analysis Facility (IRAF\footnote{http://iraf.net})  was used to perform aperture photometry to get the flux of the extraordinary ($I_{e}$) and ordinary ($I_{o}$) images of each star in the CCD frame. The ratio $R(\alpha)$ at each position of HWP is given by the equation
 
\begin{center}
\begin{equation}
    R\left(\alpha\right) =   \frac{\frac{I_{e}}{I_{o}}-1}{\frac{I_{e}}{I_{o}} + 1} = P  \, \cos \left(2\theta - 4\alpha\right)
    \label{eq:equation1}
\end{equation}
\end{center}

\noindent
where $P$ is the fraction of total linearly polarized light and $\theta$ is the polarization position angle. The error associated with $R(\alpha)$ is given by the equation

\begin{equation}
   \sigma_{R\left(\alpha\right)} = \frac{\sqrt{I_{e} + I_{o}  + 2I_{b}}}{{I_{e} + I_{o}}}  
   \label{eq:equation2}
\end{equation}

\noindent
where $I_{b}$ is average background counts around the extraordinary and ordinary images. As the responses of the CCD to the two orthogonal polarization components may not be the same and are a function of position on its surface, the actual measured signal in the two images may differ from by a factor $F$ as given in

\begin{equation}
    F = \sqrt[4]{\frac{I_{o}\left(0\right)}{I_{e}\left(45\right)} \times \frac{I_{o}\left(45\right)}{I_{e}\left(0\right)} \times \frac{I_{o}\left(22.5\right)}{I_{e}\left(67.5\right)} \times \frac{I_{o}\left(67.5\right)}{I_{e}\left(22.5\right)}}
    \label{eq:equation3}
\end{equation}

\noindent
Now, the ratio $R(\alpha)$ is given by

\begin{equation}
    R\left(\alpha\right)=\frac{\frac{I_{e}}{I_{o}} \times F-1}{\frac{I_{e}}{I_{o}} \times F+1}
    \label{eq:equation4}
\end{equation}

Correction for instrumental polarization and zero-point calibration of position angle were done using unpolarized and polarized standard stars, respectively.  The instrumental polarization was found to be less than 0.3\% in all observed passbands \citep[see also][etc.]{2007MNRAS.378..881M, 2009MNRAS.396.1004P, 2011MNRAS.411.1418E, 2015ApJ...803L..20S, 2016MNRAS.457.3178P}. 
The polarization results for polarized standard stars after the correction of instrumental polarization are given in Table \ref{tab:standard} along with the values derived by \cite{1992AJ....104.1563S}. As the overlapping of an ordinary image with the adjacent extraordinary image cannot be avoided due to the absence of a grid in the AIMPOL, to take care of target sources we have selected only those isolated stars that do not show any overlapping of the ordinary image with an adjacent extraordinary image. Furthermore, we have considered only those stars that have an error of less than 50\% in the degree of polarization and/or position angle. This turned out to be only 73 stars in the observed fov of the cluster Alessi 1. We have performed astrometry of observed sources using CCMAP and CCTRAN packages in IRAF. The rms error in the performed astrometry was around 0\arcsec.28.

\subsection{Gaia Archive}
We have also used \textit{Gaia} DR2 data \citep{2016A&A...595A...1G,2018A&A...616A...1G,2018A&A...616A...2L,2018A&A...616A...9L} to determine the basic parameters and membership of the cluster Alessi 1. We have extracted the data of the stars within a 26\arcmin\, radius of the cluster, which was also the center of the polarimetric observations. We have chosen this radius because the angular radius of the cluster was estimated to be 25.5\arcmin\, by \citet{2013A&A...558A..53K}. We have taken only those sources from \textit{Gaia} DR2  for which the three conditions for a  five parameter (positions, parallax, and proper motions) solution are met \citep[see][]{2018A&A...616A...2L}. These conditions are as follows: the mean \textit{G-}band magnitude should be  brighter than 21.0 mag, the number of distinct observation epochs should be  greater than 6, and the parameter $astrometric\_sigma5d\_max$ $\leq (1.2$ mas$) \times \gamma(G)$, where $\gamma(G) = max[1,10^{0.2(G-18)}]$.
Sources for which proper motion in R.A. ($\mu_\alpha$) and decl. ($\mu_\delta$) were outside the range of  -30 to +30  mas yr$^{-1}$ were also discarded. After applying the above conditions, a total of 11,004 sources were found in the selected region of the cluster Alessi 1. The parameters, $\mu_\alpha$, $\mu_\delta$, parallax, the magnitude in the \textit{G} band ($G$), color ($G_{BP}-G_{RP}$), and color excess $[E(G_{BP}-G_{RP})]$ of \textit{Gaia} sources were extracted from the \textit{Gaia} archive for our analysis. 

We have cross-matched our observed sources with the \textit{Gaia} sources using the CDS X-Match Service\footnote{http://cdsxmatch.u-strasbg.fr/}. We have used positional error cross-match criteria with 5$\sigma$ for the 100 \% completeness and maximum distance up to 5\arcsec. Thus, out of 73 polarimetrically observed stars, 68 stars were cross-matched with the \textit{Gaia} sources using these criteria. We found that 44 stars were matched within  1\arcsec\,(1.05$\sigma$) and 7 stars were matched within 1 - 2\arcsec\,(2.29$\sigma$), whereas in the offset ranges of 2-3\arcsec\,(3.48$\sigma$ ) and 3 - 4\arcsec\,(4.58$\sigma$), 9 and 8 stars were matched, respectively.

\section{Results, analysis, and discussion} \label{sec:analy}
The degree of polarization and position angle for 73 stars toward the cluster region in \textit{B}, \textit{V}, \textit{R}, and \textit{I} photometric passbands, along with their \textit{Gaia} IDs, are given in Table \ref{tab:pthe}. In the first column, the serial numbers for all stars are given. \textit{Gaia} ID, distances between observed stars and \textit{Gaia} positions, and distances in terms of $\sigma$ are given in columns 2-4, respectively, for all 68 cross-matched stars. The degree of polarization and corresponding position angle for all stars in four different filters are given in columns 5-12. For the remaining five stars, R.A. and decl. are listed in the table footnote, along  with their serial numbers.    

\startlongtable
\begin{deluxetable*}{llcccccccccr}
\centering
\tabletypesize{\scriptsize}
\tablecaption{Observed Degree of Polarization and Position Angle in the \textit{B}, \textit{V}, \textit{R} and \textit{I} Bands of the Stars toward the Cluster Alessi 1}
\label{tab:pthe}
\setlength\tabcolsep{4.0pt}
\tablehead{
\colhead{S. No.} & \colhead{Gaia ID} & \colhead{Offset(\arcsec)} & \colhead{$\sigma$} &  \colhead{$P_{B}\left(\%\right)$} & \colhead{$\theta_{B}\left({^o}\right)$} & \colhead{$P_{V}\left(\%\right)$} & \colhead{$\theta_{V}\left({^o}\right)$} & \colhead{$P_{R}\left(\%\right)$} & \colhead{$\theta_{R}\left({^o}\right)$} & \colhead{$P_{I}\left(\%\right)$} & \colhead{$\theta_{I}\left({^o}\right)$}\\
 \colhead{(1)} & \colhead{(2)} & \colhead{(3)} & \colhead{(4)} & \colhead{(5)} & \colhead{(6)} & \colhead{(7)} & \colhead{(8)} & \colhead{(9)} & \colhead{(10)} &  \colhead{(11)} & \colhead{(12)}
}

\startdata
1 &   414516605226812288 & 0.89  &  1.05       & 0.79$\pm$0.08&  86.3$\pm$2.7   &  1.12$\pm$0.13&  87.8$\pm$3.3  &   0.82$\pm$0.14 &  89.0$\pm$4.8  &   0.75$\pm$0.05&  86.4$\pm$2.2  \\
2  &   402506987611296640 & 0.69  &  0.81      & 0.85$\pm$0.16&  71.2$\pm$5.3   &  0.83$\pm$0.01& 100.6$\pm$0.3  &   0.94$\pm$0.11 &  79.1$\pm$0.1  &   0.59$\pm$0.11&  75.5$\pm$0.1  \\
3 &   402508465080046208  & 0.32  &  0.38    & 2.28$\pm$0.79& 109.8$\pm$9.9   &  1.35$\pm$0.04&  85.1$\pm$0.9  &   1.47$\pm$0.05 &  83.3$\pm$1.1  &   1.16$\pm$0.46&  71.6$\pm$11.3  \\
4  &   402506575294438272  & 0.16  &  0.19     & 2.57$\pm$0.01&  92.5$\pm$0.1   &  1.02$\pm$0.02&  82.5$\pm$0.7  &   0.61$\pm$0.30 &  78.9$\pm$14.5 &   1.81$\pm$0.46& 112.6$\pm$7.2  \\
5  &   402508533799522176 & 0.67  &  0.79     & 1.28$\pm$0.05&  92.9$\pm$1.3   &  1.82$\pm$0.01&  76.0$\pm$0.1  &   - &  - &    - & - \\
6  &   402508739958066048 & 0.83  &  0.97     & 0.86$\pm$0.03& 82.3$\pm$1.1   &  0.86$\pm$0.12&  	76.7$\pm$4.6  &  0.69$\pm$0.02 &  88.0$\pm$0.8  &   1.31$\pm$0.07& 80.4 $\pm$ 1.1 	 \\
7* &   402508185904098688 & 0.04  &  0.05     & 0.83$\pm$0.05&  77.6$\pm$1.7   &  0.81$\pm$0.04&  77.1$\pm$1.5  &   0.71$\pm$0.07 &  76.3$\pm$3.0  &   0.65$\pm$0.06&  90.6$\pm$2.7  \\
8  &   402505888099785344  & 0.30  &  0.35    & 1.92$\pm$0.03&  74.6$\pm$0.5   &  0.81$\pm$0.05&  76.9$\pm$1.7  &   1.01$\pm$0.23 &  98.7$\pm$6.4  &   1.03$\pm$0.15& 101.9$\pm$4.3  \\
9* &   402505819380310016  & 0.12  &  0.14     & 0.58$\pm$0.12&  90.4$\pm$6.0   &  0.77$\pm$0.11&  82.8$\pm$0.2  &   0.95$\pm$0.11 &  84.6$\pm$0.1  &   0.63$\pm$0.14&  88.2$\pm$6.5  \\
10* &   402507331208797056 & 0.10  &  0.12      & 0.97$\pm$0.38&  78.6$\pm$11.3  &  1.07$\pm$0.05&  75.3$\pm$1.3  &   - &  - &   1.21$\pm$0.49&  77.3$\pm$11.6  \\
11 &   402507223831424768 & 0.25  &  0.29      & 0.77$\pm$0.12 &  87.0$\pm$4.6 &  0.94$\pm$0.09 &	76.7$\pm$2.4 & 0.67$\pm$0.09 &	76.3$\pm$2.9 &	0.77$\pm$0.15 &	82.0$\pm$8.2 \\
12* &   402507911026190592 & 0.15  &  0.18      & 0.66$\pm$0.01&  86.0$\pm$0.4   &  0.44$\pm$0.03&  89.3$\pm$2.5  &   0.98$\pm$0.02 &  80.1$\pm$0.6  &   0.62$\pm$0.03&  97.6$\pm$1.4  \\
13* &   402507537367224064  & 0.26  &  0.31      & 0.70$\pm$0.05&  88.7$\pm$2.0   &  0.83$\pm$0.06&  81.6$\pm$2.0  &   0.96$\pm$0.12 &  78.6$\pm$0.6  &   0.68$\pm$0.08&  83.3$\pm$3.6  \\
14* &   402507842306714112  & 0.16   &  0.19     & 0.61$\pm$0.10&  89.7$\pm$4.7   &  0.88$\pm$0.06&  75.6$\pm$2.1  &   0.82$\pm$0.01 &  84.9$\pm$0.1  &   0.50$\pm$0.08&  82.7$\pm$4.9  \\
15 &   414519079128682240  & 2.05  &  2.41     & 2.48$\pm$0.71& 112.8$\pm$8.3   &  0.78$\pm$0.39& 105.1$\pm$14.6 &   0.72$\pm$0.34 &  85.0$\pm$13.4 &   0.87$\pm$0.08& 105.6$\pm$2.9  \\
16  &   414894184393561088  & 0.01  &  0.02    & 0.42$\pm$0.01&  88.2$\pm$0.5   &  0.56$\pm$0.21&  87.6$\pm$10.4 &   0.51$\pm$0.01 &  78.2$\pm$0.8  &   0.51$\pm$0.01&  92.8$\pm$0.1  \\
17 &   414518980347187840  & 3.91  &  4.58     & 0.66$\pm$0.03&  90.6$\pm$1.6   &  0.80$\pm$0.03&  85.7$\pm$1.3  &   1.14$\pm$0.14 &  80.3$\pm$3.7  &   - & -\\ 
18 & -  &	- & -   & 0.98$\pm$0.31&  89.5$\pm$8.8   &  1.11$\pm$0.30&  85.1$\pm$7.5  &   1.11$\pm$0.13 &  88.7$\pm$3.3  &   0.83$\pm$0.05&  84.8$\pm$1.9  \\
19 &   414518327512167552  & 3.84  &  4.50    & 2.33$\pm$0.01&  84.3$\pm$0.1   &  0.95$\pm$0.11&  89.0$\pm$3.2  &   0.71$\pm$0.24 & 108.0$\pm$9.7  &   1.19$\pm$0.11&  92.6$\pm$2.8  \\
20  &   414518739829026944 & 2.58  &  3.03   & 3.58$\pm$0.69&  77.1$\pm$5.5   &  2.30$\pm$0.06&  72.7$\pm$0.7  &   2.49$\pm$0.26 &  83.6$\pm$3.0  &   4.46$\pm$1.83&  80.9$\pm$12.0 \\
21*  &   414518877267971456  & 0.89  &  1.05    & 0.47$\pm$0.21&  73.9$\pm$12.5  &  0.65$\pm$0.21&  77.3$\pm$9.3  &   0.95$\pm$0.06 &  73.6$\pm$1.7  &   - &  -  \\
22 &   414518808548492800 & 0.05  &  0.06      & 0.63$\pm$0.08&  98.8$\pm$3.9   &  0.75$\pm$0.02&  81.1$\pm$0.9  &   1.05$\pm$0.03 &  84.1$\pm$0.8  &   0.70$\pm$0.10&  85.1$\pm$4.3  \\
23 &   414894154331125248 & 0.84  &  0.98      & 1.05$\pm$0.03& 100.8$\pm$0.9   &  1.14$\pm$0.31&  91.3$\pm$7.5  &   0.99$\pm$0.15 &  81.7$\pm$4.5  &   0.60$\pm$0.12&  86.4$\pm$5.6  \\
24 &   402884601136206208 & 3.53  &  4.13      & 0.26$\pm$0.02& 107.1$\pm$2.3   &  0.34$\pm$0.02&  77.7$\pm$2.2  &   1.66$\pm$0.24 &  85.1$\pm$4.1  &   0.97$\pm$0.01&  72.2$\pm$0.4 \\
25 &   402884596838415360  & 3.01  &  3.53     & -&-  &  0.85$\pm$0.03&  94.4$\pm$1.0  &   0.51$\pm$0.11 &  97.1$\pm$0.1  &   0.70$\pm$0.25& 103.3$\pm$10.1  \\
26 &   402509113617033088 & 0.17  &  0.20     & 0.62$\pm$0.11&  69.5$\pm$0.7   &  0.53$\pm$0.11&  91.0$\pm$0.1  &   0.61$\pm$0.20 &  82.6$\pm$9.4  &   0.72$\pm$0.10&  86.8$\pm$4.1  \\
27 & -  &- & -  & -&-   &  0.42$\pm$0.01&  61.0$\pm$0.3  &   1.44$\pm$0.67 &  79.0$\pm$13.2 &   2.26$\pm$0.05&  80.6$\pm$0.6  \\
28 &   402884360618046976 &  3.82  &  4.48      & 1.02$\pm$0.14&  93.2$\pm$4.1   &  1.63$\pm$0.29&  78.0$\pm$5.0  &   1.35$\pm$0.14 &  75.9$\pm$2.9  &   1.29$\pm$0.08&  86.0$\pm$1.8  \\
29 &   402884360618048000  &  3.79  &  4.45   & 0.48$\pm$0.11&  61.3$\pm$0.5   &  0.62$\pm$0.11&  81.5$\pm$0.3  &   0.86$\pm$0.03 &  72.2$\pm$1.0  &   0.68$\pm$0.04&  84.8$\pm$2.0  \\
30  &   402884218880832640 & 2.09  &  2.45     & 0.55$\pm$0.05&  91.6$\pm$2.9   &  0.64$\pm$0.04&  83.6$\pm$1.8  &   0.43$\pm$0.02 &  85.7$\pm$1.5  &   0.64$\pm$0.13&  91.3$\pm$6.1  \\
31 &   402508838740252288 & 1.47  &  1.72      & - & -  &  1.00$\pm$0.38&  82.9$\pm$10.6 &   0.91$\pm$0.16 & 108.4$\pm$5.0  &   1.45$\pm$0.36&  79.5$\pm$7.2  \\
32 &   402884257538834176 & 2.88  &  3.37     & - & -  &  0.63$\pm$0.01&  84.7$\pm$0.5  &  - & - &   1.30$\pm$0.01&  88.3$\pm$0.1  \\
33 & -  & 	- & -   & 1.67$\pm$0.49& 111.4$\pm$8.4   &  0.80$\pm$0.17&  93.2$\pm$6.2  &   1.20$\pm$0.16 &  68.6$\pm$3.9  &   - & - \\
34 &   414520835773063808 & 1.51  &  1.77      & 0.54$\pm$0.05&  72.8$\pm$3.1   &  1.07$\pm$0.10&  74.0$\pm$2.8  &   0.85$\pm$0.15 &  79.0$\pm$4.9  &   0.51$\pm$0.15&  75.9$\pm$8.8  \\
35 &   414520423456212864 & 0.29  &  0.34      & 0.84$\pm$0.09&  87.3$\pm$3.0   &  0.91$\pm$0.06&  88.3$\pm$2.0  &   0.57$\pm$0.11 &  85.4$\pm$0.4  &   0.56$\pm$0.09&  77.0$\pm$4.7  \\
36 &   414520320376999936 & 1.90 &  2.23     & 0.96$\pm$0.35&  89.8$\pm$10.5  &  1.21$\pm$0.31&  77.8$\pm$7.3  &   0.90$\pm$0.07 &  71.9$\pm$2.3  &   1.56$\pm$0.28&  66.3$\pm$5.2  \\
37 &   414518018274513920 & 3.75  &  4.39     & 1.33$\pm$0.04&  95.1$\pm$0.8   &  0.92$\pm$0.02&  86.5$\pm$0.6  &   0.94$\pm$0.19 &  76.0$\pm$5.7  &   0.48$\pm$0.04&  93.4$\pm$2.5  \\
38 &   414517846475834112  & 1.15  &  1.35     & - & -  &  0.51$\pm$0.11&  72.8$\pm$0.1  &   1.49$\pm$0.26 &  72.4$\pm$5.1  &   1.28$\pm$0.11& 113.8$\pm$0.1 \\
39 &   414517812116098816 & 0.13  &  0.15      & 0.62$\pm$0.07&  86.5$\pm$3.5   &  1.03$\pm$0.02&  85.2$\pm$0.7  &   0.79$\pm$0.12 &  84.0$\pm$1.0  &   0.72$\pm$0.06&  87.7$\pm$2.7  \\ 
40 &    414517228000553984 & 1.57  &  1.85     & - & - & -  &-&   1.12$\pm$0.32 &  53.7$\pm$8.0  &   0.98$\pm$0.21&  98.3$\pm$6.3 \\ 
41 &   414517605957672064 & 0.06  &  0.07      & 0.56$\pm$0.13&  89.0$\pm$6.9   &  0.70$\pm$0.01&  86.9$\pm$0.2  &   1.00$\pm$0.17 &  63.5$\pm$5.0  &   0.62$\pm$0.09&  87.7$\pm$4.4  \\
42 &   414516811385241600  & 0.26  &  0.31      & 0.69$\pm$0.10&  89.6$\pm$4.4   &  0.84$\pm$0.04&  86.0$\pm$1.3  &   0.63$\pm$0.07 &  80.5$\pm$3.2  &   0.61$\pm$0.14&  87.2$\pm$6.8  \\
43 &   414519186505623936 & 3.83   &  4.49      & 1.87$\pm$0.02& 103.4$\pm$0.3   &  0.93$\pm$0.01&  79.3$\pm$0.2  &   0.93$\pm$0.09 &  74.3$\pm$2.7  &   0.92$\pm$0.01&  93.1$\pm$0.2  \\
44 &   414518430591386496  & 2.59  &  3.04      &  - &  - &  0.80$\pm$0.01&  87.7$\pm$0.4  &   0.60$\pm$0.11 &  76.5$\pm$0.3  &   0.56$\pm$0.07&  81.7$\pm$3.8  \\
45 &   414516300287631616 & 0.29  &  0.34     & 1.30$\pm$0.05&  96.9$\pm$1.2   &  0.66$\pm$0.17&  78.4$\pm$7.5  &   0.68$\pm$0.04 &  73.8$\pm$1.6  &   0.88$\pm$0.26&  93.8$\pm$8.6  \\
46 &  - & - &-  &  0.65$\pm$0.01&  88.2$\pm$0.1   &  0.40$\pm$0.05&  77.2$\pm$4.0  &   0.55$\pm$0.03 &  68.1$\pm$1.8  &   0.77$\pm$0.07&  85.7$\pm$2.7  \\
47* &   414516437726588800  & 0.05  &  0.06     & 0.66$\pm$0.11&  87.9$\pm$0.1   &  0.45$\pm$0.13&  67.3$\pm$8.5  &   0.47$\pm$0.11 &  72.4$\pm$0.3  &   0.71$\pm$0.18&  83.0$\pm$7.5  \\
48* &   402506369136008832 & 0.06   &  0.08    & 0.69$\pm$0.05&  95.6$\pm$2.1   &  -  &-   &  0.69$\pm$0.06 & 67.3$\pm$2.6  &  0.86$\pm$0.15&  82.1$\pm$5.2\\
49* &   402506162977579648 & 0.05  &  0.06      & 0.85$\pm$0.04 & 98.1$\pm$1.4	& - &- &	0.82$\pm$0.05 &	78.5$\pm$1.9 &	1.02$\pm$0.15	& 91.8$\pm$3.6 \\
50 &   402506128617841920 & 0.22   &  0.26       & 0.82  $\pm$0.15 &  84.7$\pm$6.3	&  0.69$\pm$0.05 &	84.4$\pm$0.2 &	0.71$\pm$0.07 &	70.1$\pm$1.2 &	0.99$\pm$0.07	 & 81.5$\pm$2.1 \\
51* &   402505991178890752 & 0.15   &  0.17      & 0.86$\pm$0.18&  88.0$\pm$6.2   &  0.65$\pm$0.03&  73.6$\pm$1.5  &   0.65$\pm$0.11 &  67.8$\pm$4.9  &   0.89$\pm$0.15&  88.2$\pm$5.0  \\
52 &   402505991178891136  & 0.16   &  0.19     & 1.15$\pm$0.13 &  86.4$\pm$3.2  & 0.54$\pm$0.07 &	83.9$\pm$2.1  &   -  &  -  & 0.96$\pm$0.08 & 73.8$\pm$2.4   \\
53* &   402500317524014976 & 0.15   &  0.18      & 1.01$\pm$0.14 &  87.1$\pm$4.2	&  0.70$\pm$0.07 &	75.2$\pm$3.1 &	0.50$\pm$0.10 &	78.3$\pm$7.4 &	1.30$\pm$0.17 &	80.5$\pm$4.1 \\
54 &   402500390541534848 & 0.35  &  0.41     & 0.79$\pm$0.07&  93.2$\pm$2.8   &  0.64$\pm$0.03&  73.6$\pm$1.5  &   0.80$\pm$0.11 &  71.1$\pm$0.1  &   0.92$\pm$0.05&  90.9$\pm$1.7  \\
55 &   414515887970782592  & 0.08  &  0.10     & - &  -  &  0.83$\pm$0.10&  61.3$\pm$3.5  &   1.14$\pm$0.37 &  83.0$\pm$9.4  &   0.92$\pm$0.05&  90.9$\pm$1.7  \\
56 &   414516158550216320  & 0.07  &  0.09     & 0.95$\pm$0.01&  83.3$\pm$0.3   &  0.94$\pm$0.01&  77.0$\pm$0.5  &   0.86$\pm$0.05 &  62.9$\pm$1.8  &   1.22$\pm$0.36&  77.1$\pm$8.5  \\
57* &   414509978093147136 & 0.16  &  0.19      & 1.04$\pm$0.02&  81.0$\pm$0.7   &  0.79$\pm$0.03&  72.2$\pm$1.3  &   0.87$\pm$0.11 &  72.3$\pm$0.4  &   1.08$\pm$0.19&  79.1$\pm$5.1  \\
58 &   414510149894478080  & 0.08  &  0.09     & 1.36$\pm$0.11&  77.1$\pm$0.1   &  0.86$\pm$0.05&  72.1$\pm$1.9  &   1.01$\pm$0.16 &  72.2$\pm$4.6  &   0.88$\pm$0.29& 114.6$\pm$9.6 \\
59 &  -  & 	- & -    & 1.00$\pm$0.38&  89.3$\pm$11.0  &  0.61$\pm$0.07&  84.4$\pm$3.6  &   0.74$\pm$0.02 &  76.8$\pm$1.0  &   0.97$\pm$0.13&  81.3$\pm$4.0  \\
60 &   402499291029914752  & 2.76  &  3.24      & 1.06$\pm$0.25&  84.7$\pm$6.9   &  0.86$\pm$0.02&  64.2$\pm$0.7  &   0.86$\pm$0.07 &  73.5$\pm$2.6  &   0.93$\pm$0.19&  73.3$\pm$6.0  \\
61 &   402499256670177536 & 2.97  &  3.48      & 2.01$\pm$0.48&  73.4$\pm$6.8   &  0.86$\pm$0.29&  96.4$\pm$9.8  &   -  & -  &   1.45$\pm$0.15 &  79.5$\pm$3.1 \\
62 &   402500012584419840 & 2.04  &  2.39      & 1.36$\pm$0.35&  95.4$\pm$7.5   &  0.95$\pm$0.13&  69.5$\pm$4.0  &   0.80$\pm$0.03 &  64.0$\pm$1.3  &   1.11$\pm$0.15 &  81.1$\pm$3.9  \\
63 &   402500012584421120 & 1.95  &  2.29      & 1.52$\pm$0.10& 109.3$\pm$2.0   &  0.87$\pm$0.23&  88.4$\pm$7.6  &   0.63$\pm$0.06 & 120.0$\pm$2.7  &   1.11$\pm$0.23&   84.0$\pm$6.0  \\
64 &   402505334045815680 & 2.29 &  2.69      & 0.79$\pm$0.14&  91.2$\pm$5.2   &  0.67$\pm$0.11&  89.9$\pm$4.7  &   -  &-  &   0.45$\pm$0.01&   93.6$\pm$1.0 \\
65 &   402506059898367104 & 1.55  &  1.82      & - &  -  &  0.36$\pm$0.04&  66.0$\pm$3.1  &   1.29$\pm$0.20 &  77.0$\pm$4.4  &   0.80$\pm$0.28&   81.8$\pm$10.2 \\
66 &   402507670508022784  & 0.14 	  &  0.16      & -& -   &  0.88$\pm$0.24&  81.6$\pm$7.8  &   0.88$\pm$0.16 &  76.8$\pm$5.3  &   0.86$\pm$0.01&  87.4$\pm$0.1  \\
67* &   402507571726963968 &  0.19  &  0.22    & - &  -  &  0.64$\pm$0.03&  75.3$\pm$1.3  &   0.75$\pm$0.10 &  78.9$\pm$3.8  &   0.91$\pm$0.02&  88.0$\pm$0.8 \\
68 &   402504788588155264 & 0.11  &  0.13     & 0.48$\pm$0.07& 147.8$\pm$4.4   &  0.52$\pm$0.18& 101.4$\pm$10.1 &   0.25$\pm$0.08 &  89.9$\pm$8.8  &   0.89$\pm$0.26&  84.9$\pm$8.4  \\
69  &   402879825133045760 & 0.20  &  0.23     & 0.12$\pm$0.01& 139.5$\pm$3.3   &  0.36$\pm$0.02&  58.5$\pm$1.8  &   0.89$\pm$0.08 &  73.6$\pm$2.6  &   0.66$\pm$0.15&  91.5$\pm$6.6  \\
70 &   402504754228418688 & 0.11  &  0.13       & 1.45$\pm$0.46& 116.2$\pm$9.6   &  0.89$\pm$0.21&  77.6$\pm$7.0  &   1.37$\pm$0.02 &  69.9$\pm$0.4  &   2.16$\pm$0.20&  84.2$\pm$2.6  \\
71* &   402880684126058880 & 0.07  &  0.08      & 0.53$\pm$0.07& 101.4$\pm$3.7   &  -  & -  &  -  &  -  &   0.87$\pm$0.12&  85.1$\pm$4.0 \\
72 &   402883291167896832 & 0.27  &  0.31       & 0.80$\pm$0.11&  93.4$\pm$3.9   &  0.25$\pm$0.02&  52.6$\pm$3.0   &  0.77$\pm$0.11 &  83.8$\pm$4.1  &   0.94$\pm$0.02&  96.7$\pm$0.6  \\
73 &   402883428606850816 & 0.87  &  1.02     & 1.05$\pm$0.19&  72.4$\pm$5.2   &  0.85$\pm$0.08&  71.6$\pm$2.7  &   1.41$\pm$0.14 &  77.6$\pm$3.0  &   1.16$\pm$0.01&  80.0$\pm$0.3  \\
\enddata
\tablecomments{* Member stars of the cluster Alessi 1.  S.No. 18, R.A.:13.279948, decl.:49.693928. S.No. 27, R.A.:13.389086, decl.:49.724924. S.No. 33, R.A.:13.106471, decl.:49.691115. S.No. 46, R.A.:13.254734, decl.:49.581148. S.No. 59, R.A.:13.438451, decl.:49.490717.}

\end{deluxetable*}

\begin{figure}[b]
\centering
\subfigure[Histogram of degree of polarization]{\includegraphics[width=\columnwidth]{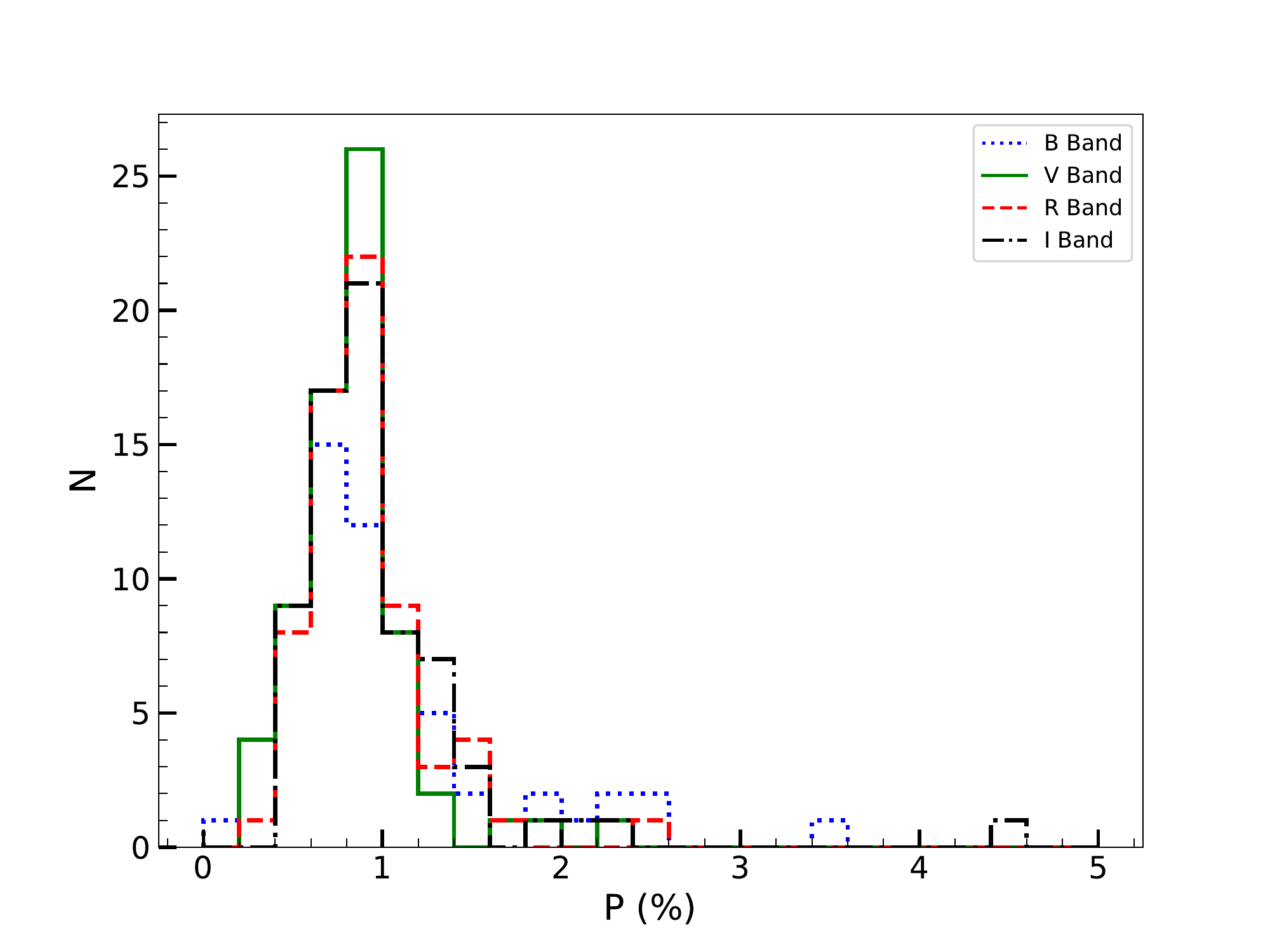} \label{fig:hist_p}}  
\subfigure[Histogram of angle of polarization]{\includegraphics[width=\columnwidth]{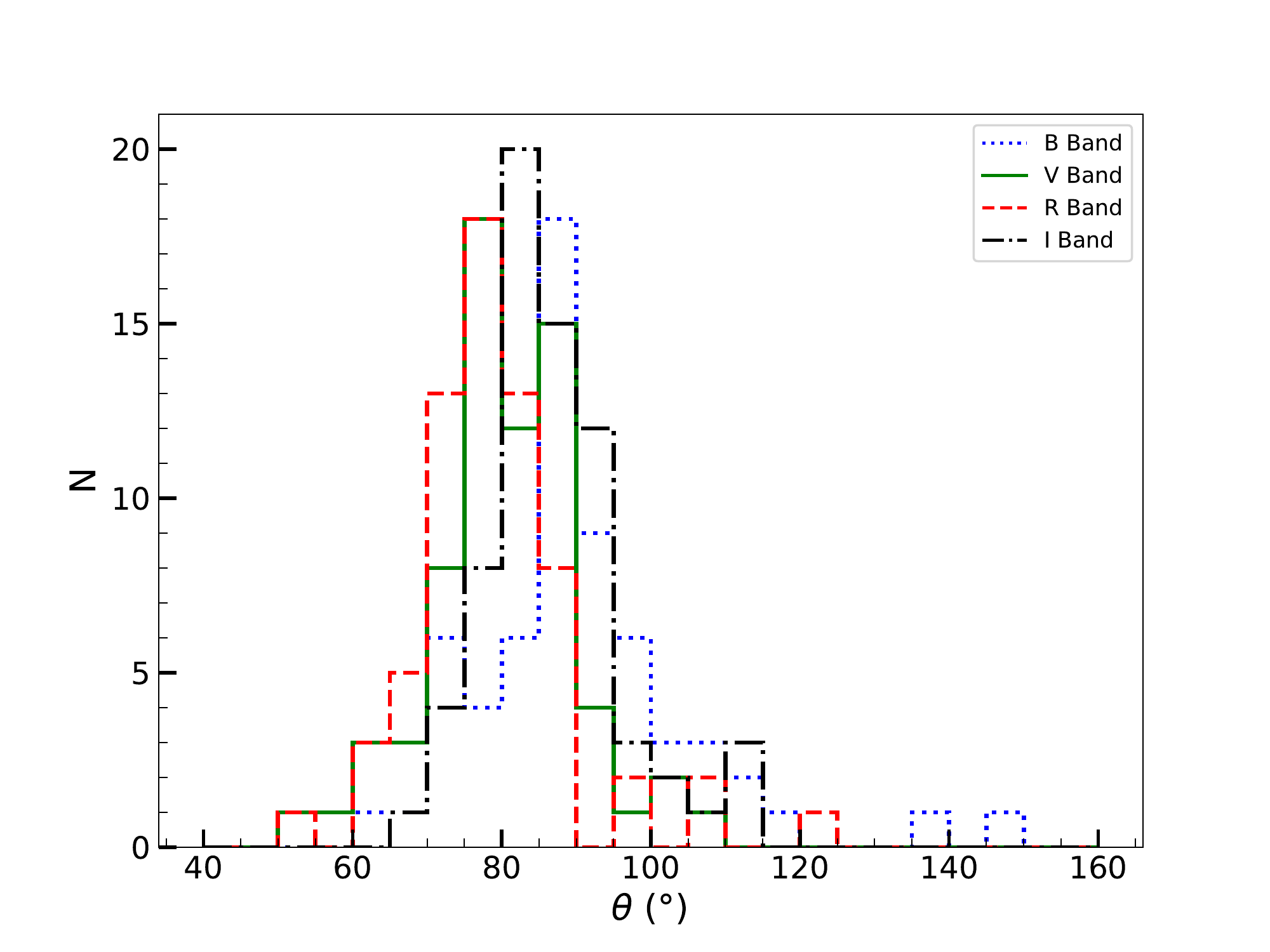}\label{fig:hist_t}}  \caption{Distribution of degree of polarization and position angle of  observed stars in the \textit{B}, \textit{V}, \textit{R}, and \textit{I} bands toward the cluster Alessi 1.}
\label{fig:hist_pt}
\end{figure}

The distributions of $P$ and $\theta$ in the \textit{B}, \textit{V}, \textit{R}, and \textit{I} bands are shown in Figure \ref{fig:hist_pt} for all observed stars. The value of $P$ was found in the range of 0.12\%-4.46\% for the \textit{B}, \textit{V}, \textit{R}, and \textit{I} bands for all stars. However,  for the majority of stars, the values of $P$ were found to be in the range of 0.4\%-1.4\% in all bands. The value of $\theta$  is distributed from $52^{o}.6$  to $147^{o}.8$ in the \textit{B}, \textit{V}, \textit{R}, and \textit{I} bands. The weighted mean values of $P$ and $\theta$ were found to be $0.77\pm0.01$\% and $84^{o}.4\pm0^{o}.1$, $0.70\pm0.01$\% and $75^{o}.2\pm0^{o}.2$, $0.76\pm0.01$\% and $74^{o}.9\pm0^{o}.1$, and $0.96\pm0.02$\% and $83^{o}.7\pm0^{o}.6$ in the \textit{B}, \textit{V}, \textit{R}, and \textit{I} bands, respectively, for member stars of the cluster Alessi 1. 

\begin{figure*}
	\centering
\includegraphics[width=1.6\columnwidth,trim={0.2cm 0.2cm 0.0cm 0.0cm}]{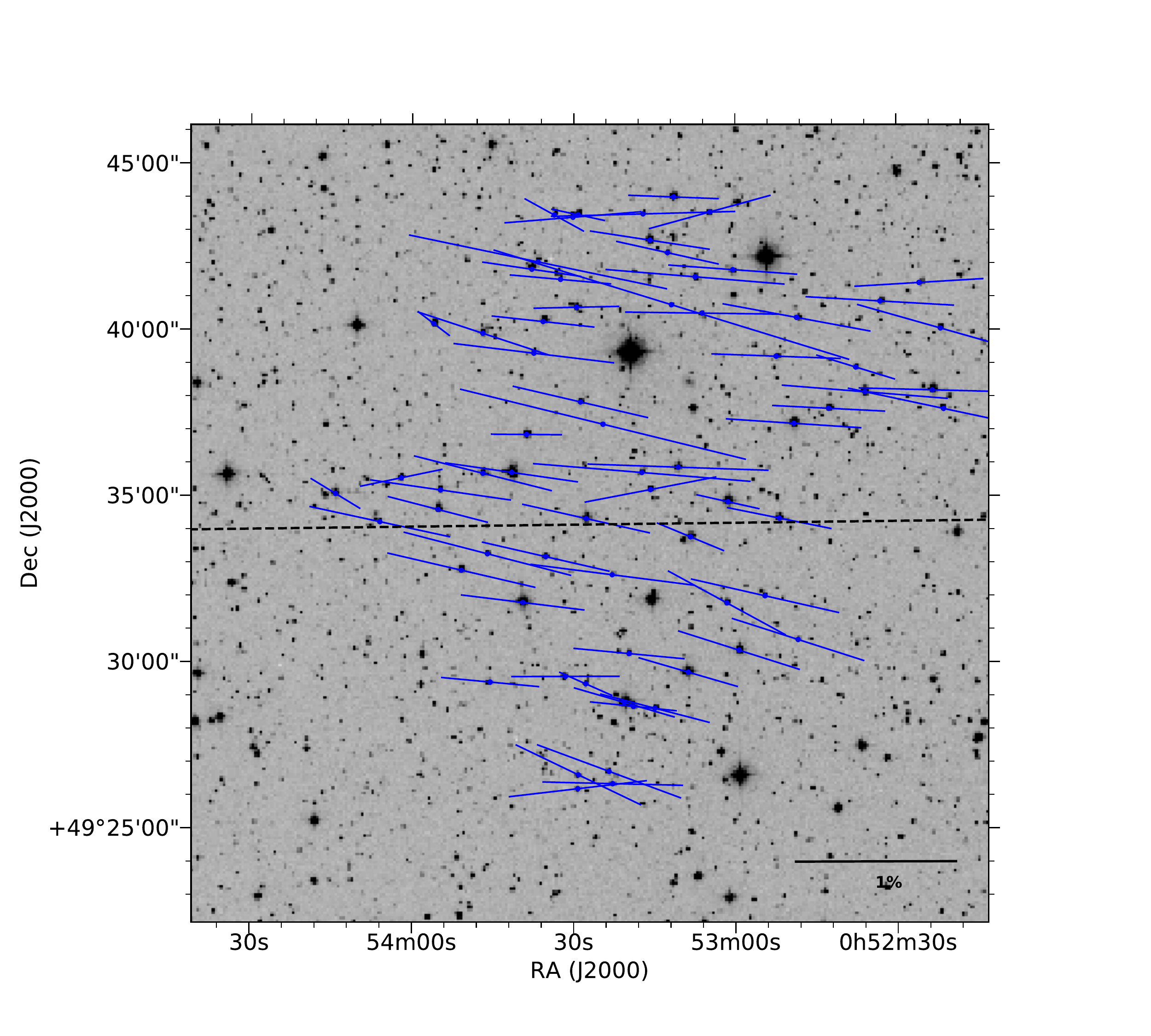}
\caption{Stellar polarization in the \textit{V} band superimposed on a 24\arcmin $\times$ 24\arcmin\, DSS image of the field containing Alessi 1. The length of the polarization vectors is proportional to $P_{V}$. A line of magnitude of 1\% polarization value is drawn for a reference. The dashed black line indicates the orientation of the projection of the Galactic plane at $b = -13^{o}.30$.}
    \label{fig:vector_plot}
\end{figure*}

We have shown the sky projection of polarization vectors in the \textit{V} band for observed stars over the digitized sky survey (DSS\footnote{http://archive.eso.org/dss/dss}) \textit{R}-band image of the cluster Alessi 1 in Figure \ref{fig:vector_plot}. The polarization vectors are drawn keeping the stars at the center. Observed stars are marked with a blue dot and the polarization vectors are drawn with blue lines. The length of polarization vectors is proportional to the percentage of polarization. A line of the magnitude of 1 \% polarization value is shown for the reference at the bottom right of Figure \ref{fig:vector_plot}. The dashed black line indicates the orientation of the projection of the Galactic plane at $b = -13^{o}.30$.  Most of the polarization vectors were found to be almost parallel to the direction of the Galactic plane.  In the majority of polarimetric studies of open clusters, it was found that the alignments of polarization vectors are almost parallel to the direction of the Galactic plane \citep[e.g.][]{,2004A&A...419..965M,2007A&A...462..621V,2007MNRAS.378..881M,2010A&A...513A..75O, 2010MNRAS.403.1577M, 2011MNRAS.411.1418E}, while toward some other lines of sight, the second component of the magnetic field was also found, which was slightly inclined to the Galactic plane \citep{2008MNRAS.388..105M,2012MNRAS.419.2587E}. \citet [][]{1951ApJ...114..206D} suggested the dominant mechanism for alignment of dust grains in the ISM  could be due to the presence of the small non-conservative torque produced by paramagnetic relaxation, and that this torque tends to align the short axis of the rapidly spinning grain along the magnetic field. However, there are some other studies where polarization vectors are found to not be aligned with the Galactic plane, indicating the dust experienced perturbation \citep{1978Ap&SS..54..425E,2010MNRAS.403.2041V, 2018RMxAA..54..293V}.  
Thus, the dust grains along the line of sight of the cluster Alessi 1 are possibly aligned by the Galactic magnetic field and are located in an unperturbed place of our Galaxy.
\\
\\

\subsection{Membership }
\label{sec:Membership} 
\subsubsection{Polarimetric Approach} \label{sec:Polarimetric Approach}
\begin{figure*}
\centering
\subfigure[$Q$ versus $U$ in B band]{\includegraphics[width=\columnwidth]{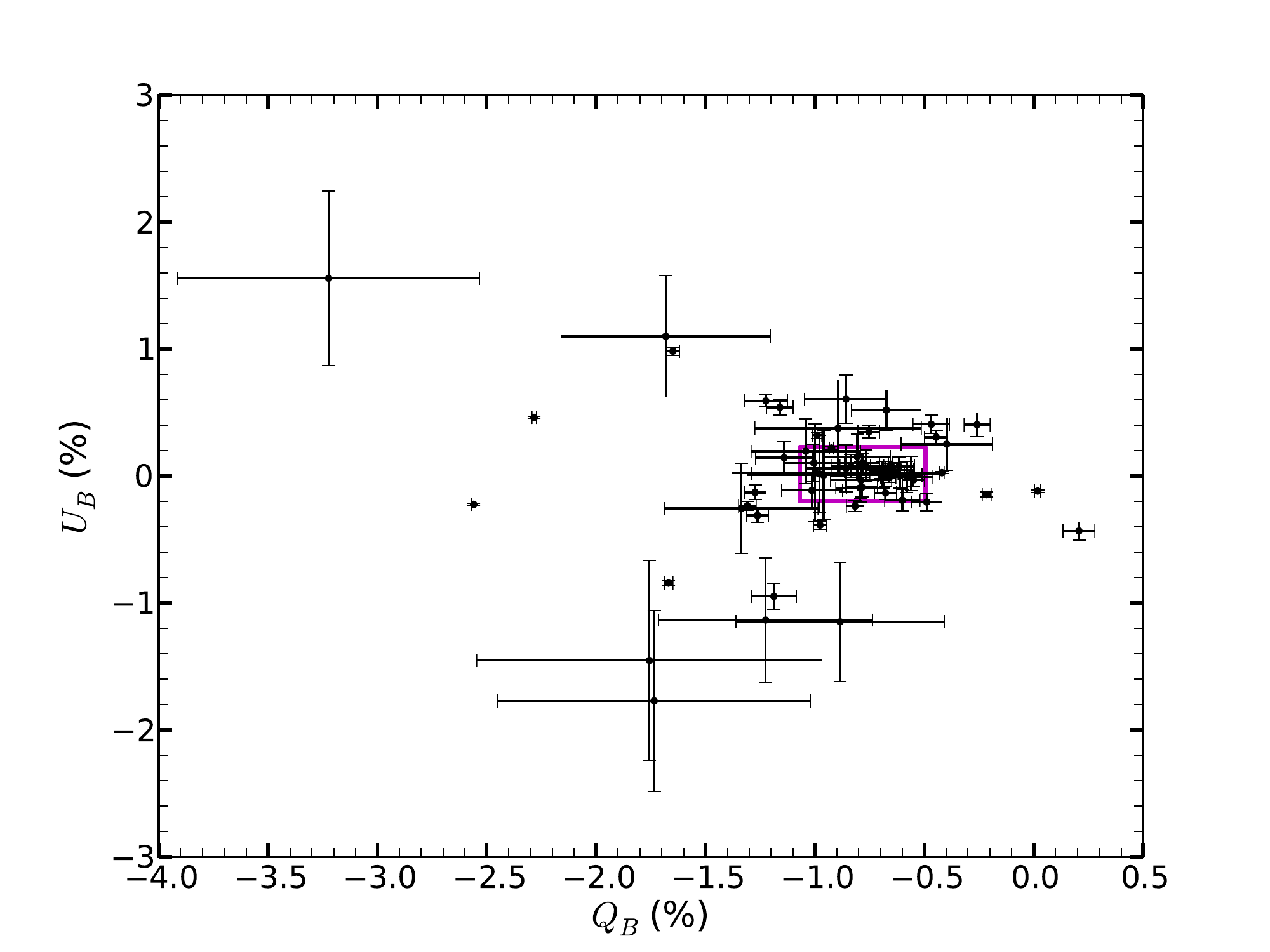}\label{fig:qub}}
\subfigure[$Q$ versus $U$ in V band]{\includegraphics[width=\columnwidth]{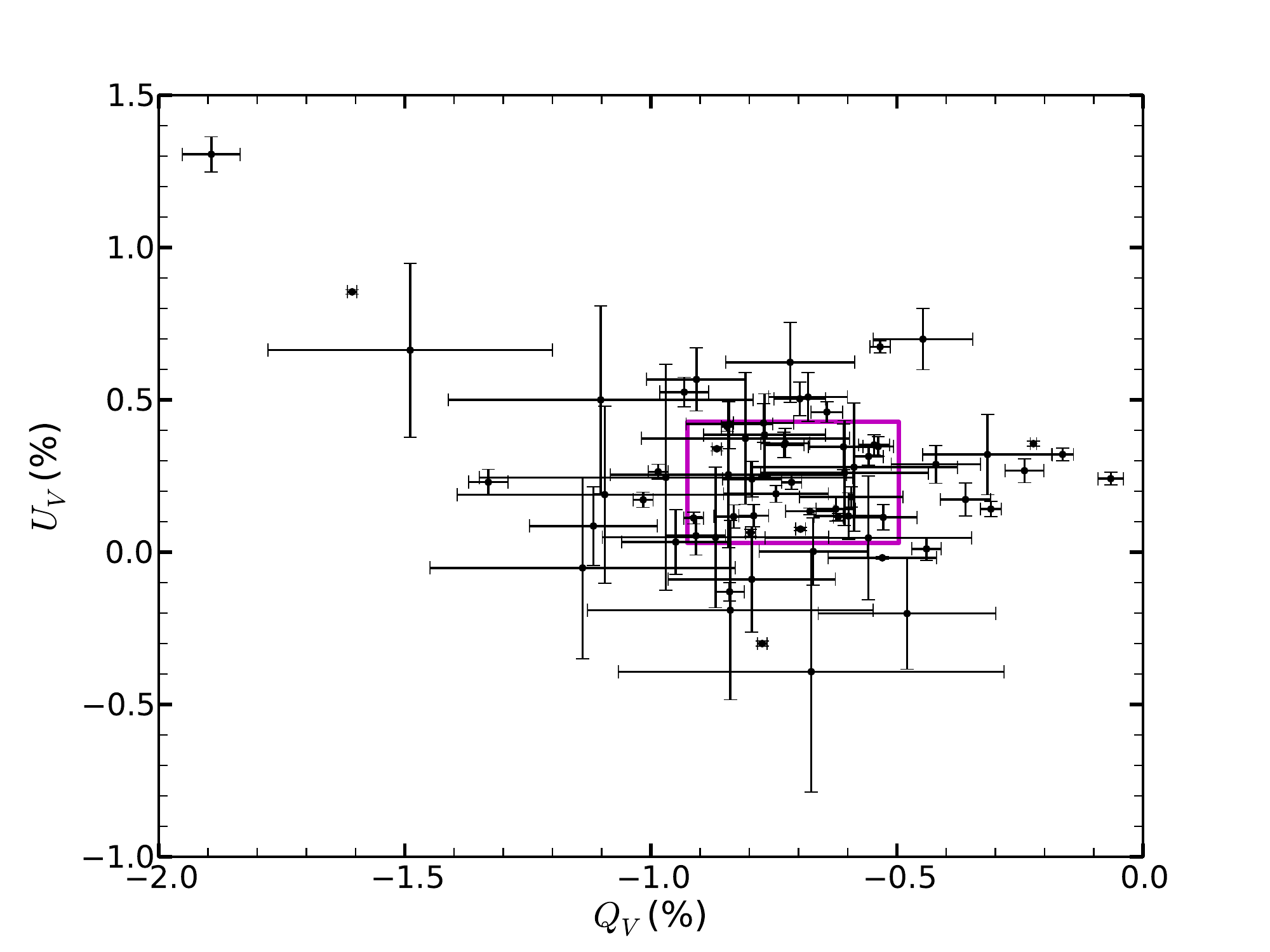}\label{fig:quv}}
\subfigure[$Q$ versus $U$ in R band]{\includegraphics[width=\columnwidth]{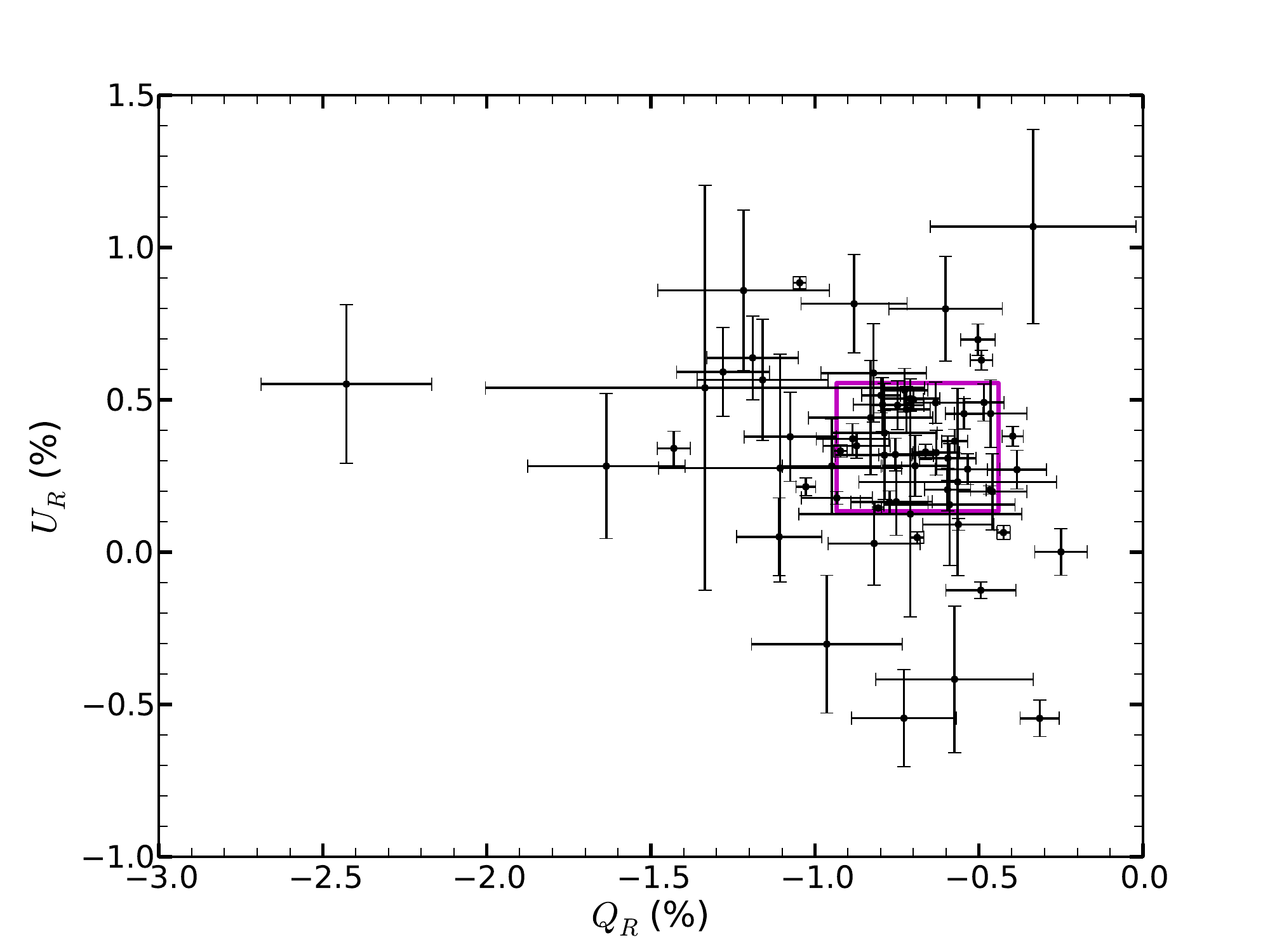}\label{fig:qur}}
\subfigure[$Q$ versus $U$ in I band]{\includegraphics[width=\columnwidth]{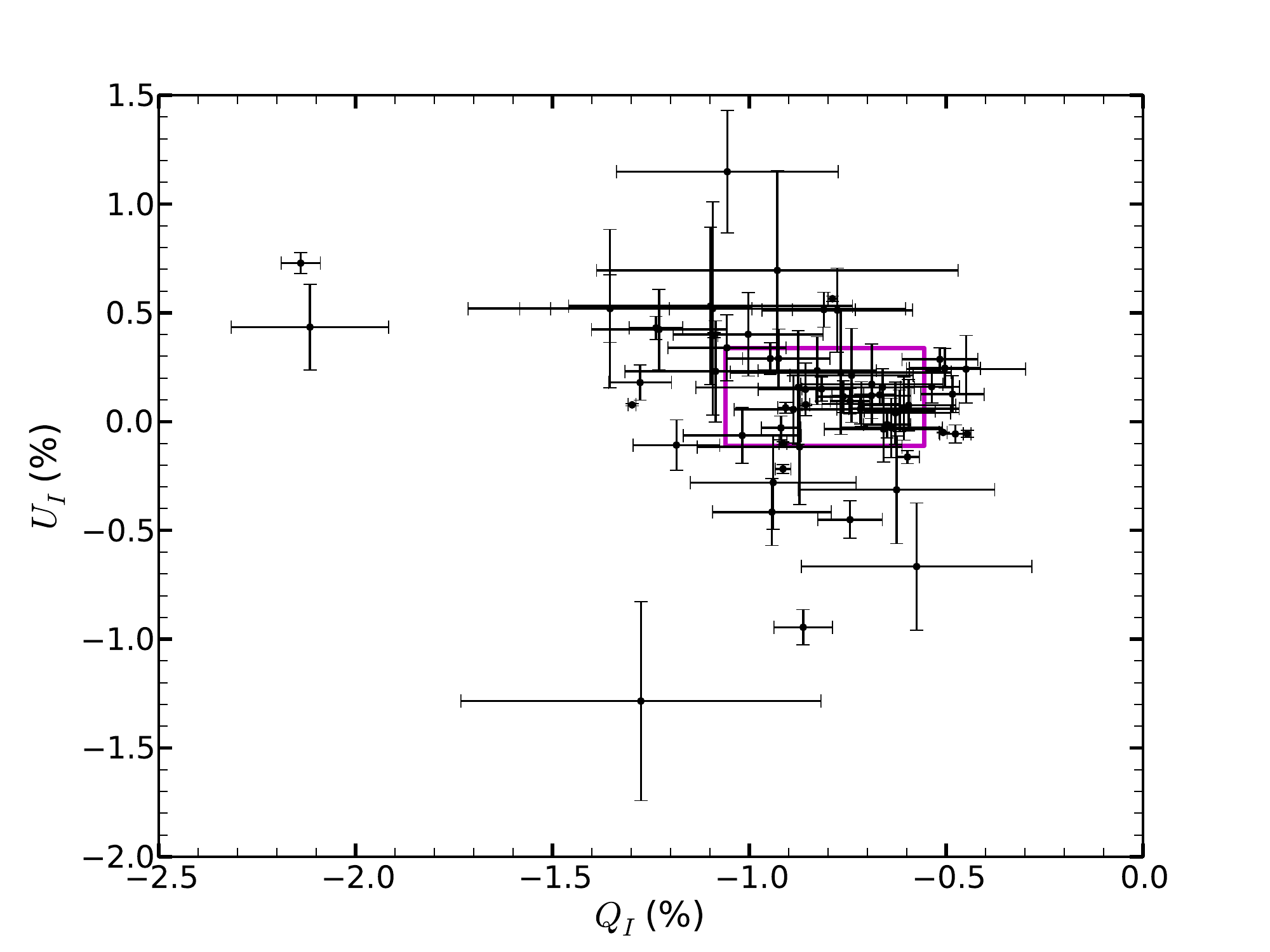}\label{fig:qui}}
\caption{Plot between Stokes parameters $Q$ and $U$ for observed stars in the \textit{B}, \textit{V}, \textit{R}, and \textit{I} bands. The rectangle in each figure shows the 1$\sigma$ box from the mean values of $Q$ and $U$. One star in the \textit{I} band is far off from the mean values of $Q$ and $U$, and hence is not shown in the $Q$-$U$ plot.}
    \label{fig:QU}
\end{figure*}

The light coming from distant stars is partially plane polarized, which is thought to be due to dust grains in the ISM. Hence, the degree of polarization of a star depends on the column density of aligned dust grains that lie in front of the star, if the star is not intrinsically polarized. So, one can infer that, for the member stars of the cluster, the degree of polarization should be the same. If there is a foreground source then the degree of polarization should be lower, as the light coming from the star will face a less column density, and if there is a background source then a higher value of polarization is expected depending upon the length of the column density of grains. Hence, a plot of the  Stokes parameters $Q (= P \cos2\theta)$ and $U (= P \sin2\theta)$, also known as a Stokes plane, is a useful tool to distinguish cluster members and non-members. The members of the cluster are expected to group together in the plot and non-members are expected to appear scattered. A plot between Stokes parameters with corresponding errors is shown in Figure \ref{fig:QU} for all observed passbands. In this plot, one star in the \textit{I} band is not shown, as this was highly scattered. In Figure \ref{fig:QU}, rectangles show the 1$\sigma$ boundary from mean values of $Q$ and $U$. The mean and standard deviation ($\sigma$) of the $Q$ and $U$ were derived by fitting the Gaussian function to their distribution. This boundary can be used to differentiate the members and non-members of the cluster. Stars inside this box can be considered as probable members of the cluster. 

Using these plots, we found 9 common stars that lie inside the 1$\sigma$ box in all four passbands. Excluding these 9 stars, 20 stars were found to be common in any three bands and 19 stars were found to be common in any two bands inside the boundary of a 1$\sigma$ box. If we consider those stars within a 1$\sigma$ box of $Q$-$U$ plane, which were common at least in any two bands, as a primary step toward separating  members, we found a total of 48 stars to be  members of the cluster Alessi 1.  However, there is a strong possibility that there are non-members among these stars either due to background population or intrinsic polarization of some stars.
So this technique cannot be used alone. \citet{2013MNRAS.430.1334M} have also used polarimetric data to estimate the membership probability of some open clusters. They have seen that the polarimetric method is inaccurate for non-members stars but this can be used to estimate the  membership probability for the known member stars with unknown membership probabilities.

\subsubsection{Astrometric Approach}
\label{sec:Astrometric Approach}
\begin{figure*}
	\centering
\includegraphics[width=1.2\columnwidth]{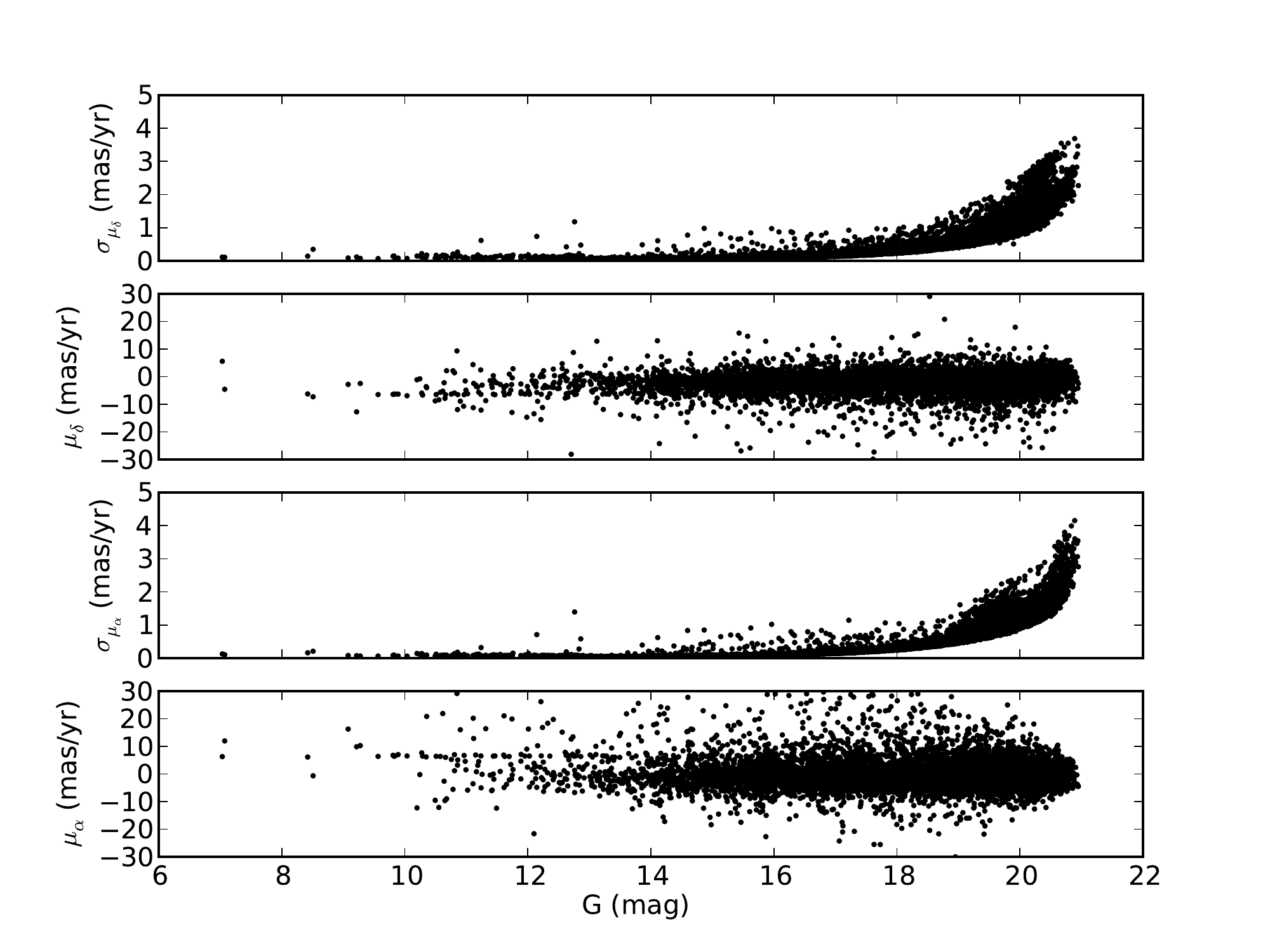}
\caption{Plot of proper motions and their errors as a function of \textit{G}-band magnitude.}
    \label{fig:pm_g}
\end{figure*}

A proper motion study is considered as a unique technique for deriving the membership information of stars in any cluster. The distributions of  $\mu_{\alpha}$,  $\mu_{\delta}$, and their corresponding errors ($\sigma_{\mu_{\alpha}}$ and $\sigma_{\mu_{\delta}}$), along with \textit{G}-band magnitude, are shown in Figure \ref{fig:pm_g}. Note that toward the fainter limit ($G > 18$), errors in both $\mu_{\alpha}$ and $\mu_{\delta}$  increase sharply and go up to 4 mas yr$^{-1}$. Therefore, for further analysis, we have taken only those sources that have both a  $\sigma_{\mu_{\alpha}}$ and $\sigma_{\mu_{\delta}}$ that are less than 1 mas yr$^{-1}$. After applying these criteria, 7598 sources out of 11,004 were left.

\begin{figure*}
	\centering
\includegraphics[width=1.4\columnwidth]{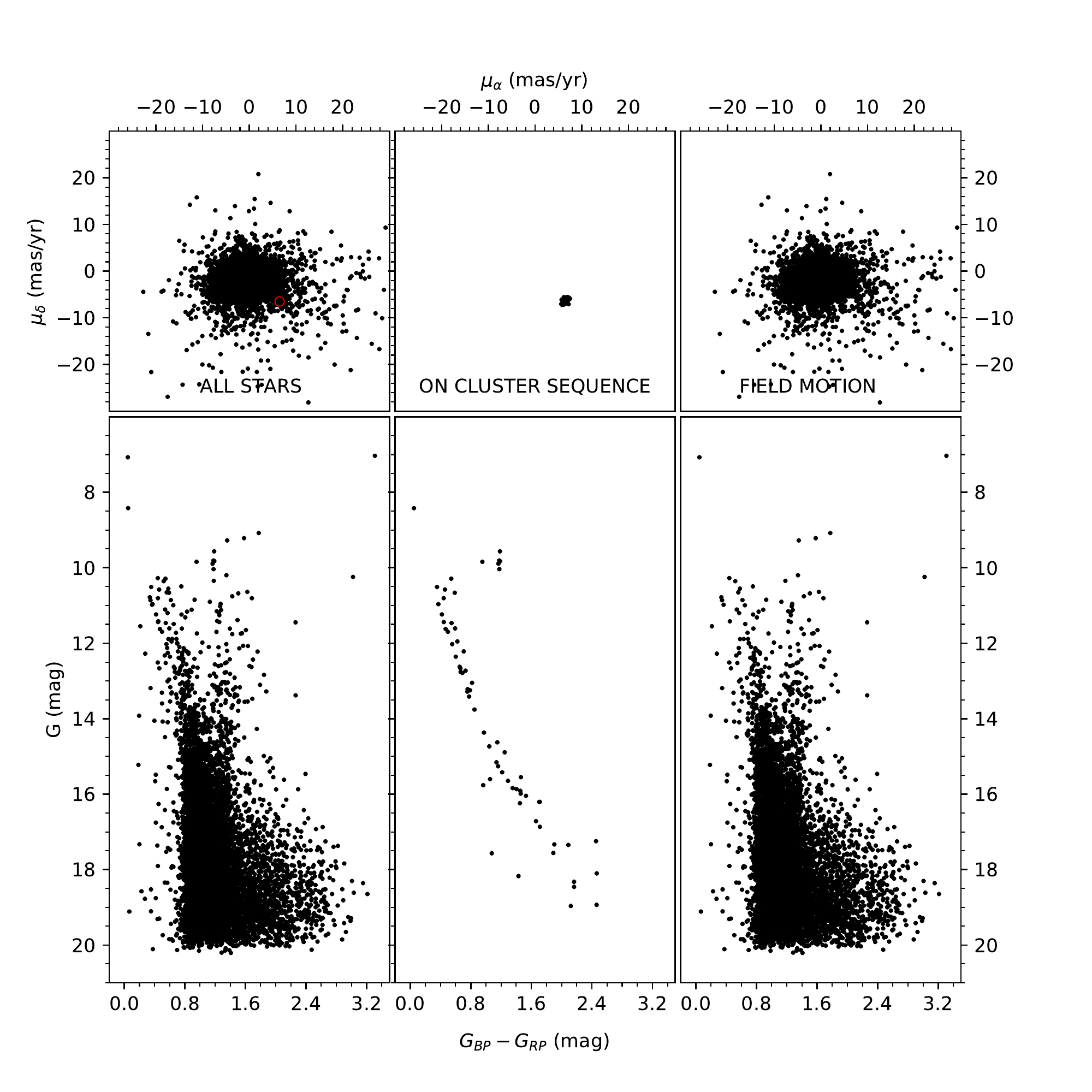}
\caption{VPD and CMD are shown for stars with proper motion errors $<$ 1 mas yr$^{-1}$, around a 26\arcmin\, radius of cluster center. A circle of 1 mas yr$^{-1}$ around the cluster centroid in the VPD defines the criterion for membership. The left, middle, and right panels show all stars, only member stars, and the field stars, respectively.}
    \label{fig:vpd_cmd}
\end{figure*}

The top left  panel of Figure \ref{fig:vpd_cmd} shows the  plot between $\mu_{\alpha}$ and $\mu_{\delta}$ [also called  vector point diagram (VPD)] of stars in the specified fov. In this plot, there is a small clustering of stars near the central coordinate ($\mu_{\alpha}$, $\mu_{\delta}$) of ($6.50$,$-6.28$) mas yr$^{-1}$. The central values of $\mu_{\alpha}$ and $\mu_{\delta}$ were estimated by fitting the Gaussian in their distribution. The bottom left panel of Figure \ref{fig:vpd_cmd} shows the corresponding CMD, where we could not identify any clear sequence of stars. To find the real cluster members, we have plotted VPD with different radii  (1 mas yr$^{-1}$ and onward) with center coordinates as mentioned above. The corresponding CMD was also plotted to identify any cluster sequence. We found that by increasing the radius of the circle beyond 1 mas yr$^{-1}$ in VPD, only the number of field stars increased, while no notable increment was seen in the main-sequence stars. So, we constrain ourselves to a 1 mas yr$^{-1}$ radius of the circle in VPD to define the membership criteria.  The chosen radius in VPD is a tradeoff between the losing cluster members and including field stars or vice versa. From this approach, we have identified 66 stars as cluster members. The middle panels of Figure \ref{fig:vpd_cmd} show the VPD and CMD for member stars, whereas the right top and bottom panels of this figure show the VPD and CMD of field stars. In the middle CMD, stars in the main sequence are well separated. Using the \textit{Gaia} DR2 data at a 25\arcmin.5 radius from the central coordinate of the cluster Alessi 1, \citet{2018A&A...618A..93C} have estimated a  ($\mu_{\alpha}$,$\mu_{\delta}$) of ($6.536$,$-6.245$) mas yr$^{-1}$ and a distance of  704.8 pc. While deriving the membership probability, they have taken only those sources for which $G \leq 18$,  typical astrometric uncertainties $<$ 0.3 mas yr$^{-1}$ in proper motion, and parallax uncertainty $\leq$ 0.15 mas.  Out of the 66 members identified by us, 40 stars have membership probability $>$ 50 \% as given by \citet{2018A&A...618A..93C}.

\subsubsection{Members in the Observed Region}
\label{sec:members_observed}
Out of 66 stars identified as members of the cluster Alessi 1, only 15 stars have been observed polarimetrically. These 15 stars share a similar degree of polarization, as they are among the 48 stars that were considered probable members of the cluster from the polarimetric approach. The small fraction of polarimetrically observed stars that we found is due to the small observed field, and does not consider the stars fainter than a \textit{G}-band magnitude of 15 due to large errors, and the locations of stars at the edge of the observed fov and/or due to overlapping of one's ordinary image with another's extraordinary image.
In Table \ref{tab:pthe},  these member stars are marked with an asterisk.  Thus, similar to the conclusion drawn by  \cite{2013MNRAS.430.1334M}, we also conclude that polarimetry alone cannot be used to determine the membership of a cluster. 

\subsection{Cluster Parameters}
\label{sec:Cluster Parameters}
\begin{figure}[h]
	\centering
\includegraphics[width=\columnwidth]{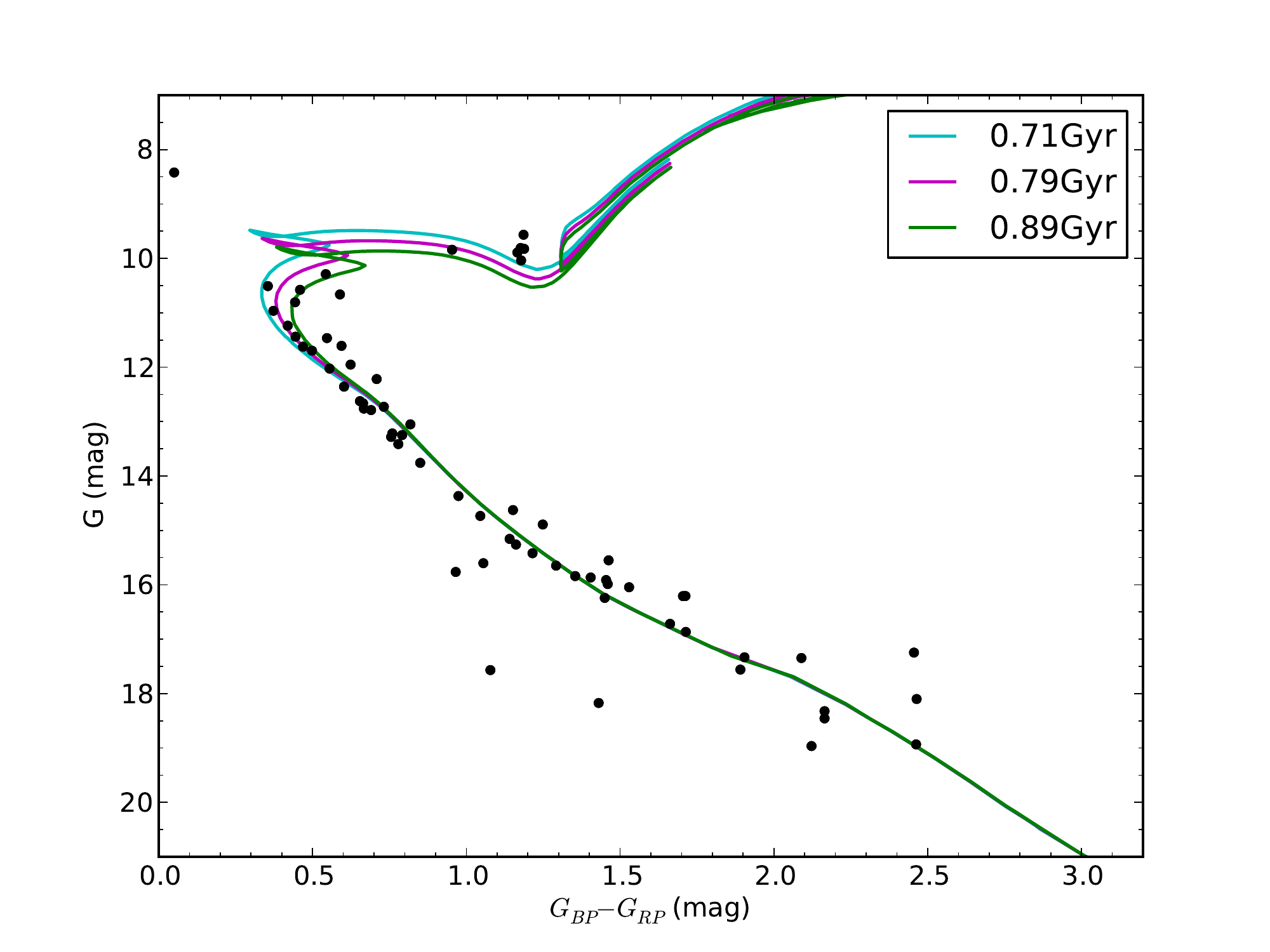}
\caption{CMD of the member stars of the cluster Alessi 1,  along with the best-matched isochrones.}
    \label{fig:isochrone}
\end{figure}

Figure \ref{fig:isochrone} shows the CMD of all 66 member stars along with the best-matched isochrones from \citet{2017ApJ...835...77M}. The isochrone with an age of 0.79 Gyr was found to be the best-matched isochrone, whereas isochrones with ages of 0.89 Gyr and 0.71 Gyr show the upper and lower limits for age estimation. 
Thus, the age of the cluster was estimated as $0.8\pm0.1$ Gyr. Using these isochrones, we have estimated the $E\left(G_{BP} - G_{RP}\right)$ and distance modulus of the cluster Alessi 1 as $0.23\pm0.05$ mag and $9.6\pm0.3$ mag, respectively. This distance modulus corresponds to a distance of  $673\pm98$ pc. While calculating the distance, the extinction in the \textit{G} band ($A_{G}$) was derived using  $A_{G} \approx 2 \times E\left(G_{BP} - G_{RP}\right)$\citep[see][]{2018A&A...616A...8A} and found to be $0.46\pm0.10$ mag. The age and distance of the cluster that we estimated by us are very close to those estimated by \citet{2004AN....325..740K, 2013A&A...558A..53K} and \citet{2019A&A...623A.108B}. 

We have also transformed  ($G_{BP}-G_{RP}$) color to ($B-V$) color using the transformation equation and coefficients of \citet{2010A&A...523A..48J}. A total of 44 stars out of 73 polarimetrically observed stars have  $E\left(G_{BP}-G_{RP}\right)$  in the \textit{Gaia} DR2. Of these, only 18 stars have $E\left(G_{BP}-G_{RP}\right)$  values with an uncertainty less than 50\%.  Therefore, for further analysis, we have considered only these 18 stars.
We  have also transformed  $E\left(G_{BP}-G_{RP}\right)$ to $E\left(B-V\right)$ and derived the relation toward that direction of the cluster Alessi 1 as  $E\left(B-V\right) = 0.7042 \times E\left(G_{BP}-G_{RP}\right) + 0.0010$. 
The average value of reddening, $E\left(B-V\right)$, was thus estimated to be  $0.16\pm0.03$ mag. Using the normal reddening law the average value of $A_{V}$ toward the cluster Alessi 1 is derived to be $0.50\pm0.09$ mag. Our derived values of $E(B-V)$ are consistent with those  given in \citet{2013A&A...558A..53K} and \cite{2019A&A...623A.108B}.

\subsection{ISM Polarization}
\label{sec:ISM Polarization}
We have checked for ISM polarization for almost all observed stars in the cluster region using the Serkowski polarization relation, which states that the  wavelength dependence of ISM polarization follows the following empirical law \citep{1975ApJ...196..261S}:

\begin{equation}
    P_{\lambda} = P_{max} \times exp[-K \times \ln^{2}\left(\lambda_{max}/\lambda\right)]
	\label{eq:equation6}
\end{equation}

\noindent
Here, $P_{\lambda}$ is the degree of polarization measured at a wavelength $\lambda$, and $\lambda_{max}$ is the wavelength at which maximum polarization $P_{max}$ occurs. The $\lambda_{max}$ is a function of the optical properties and characteristic particle size distribution of aligned grains \citep{1973ApJ...182...95C,1978ApJ...225..880M,1980ApJ...235..905W}. $K$ is the inverse of the width of the polarization curve and is taken as 1.15 for all stars \citep{1975ApJ...196..261S}. The criteria for ISM polarization are well represented by the parameter $\sigma_{1}$, which is the unit weight error of fit and quantifies the departure of our data from the Serkowski relation. If the polarization curve is well represented by a Serkowski empirical law, the value  $\sigma_{1}$ should be less than 1.6 and polarization can be considered due to an ISM origin. The value of $\sigma_{1}$  more than  1.6 and/or a small value of $\lambda_{max}$ may indicate the presence of intrinsic polarization. The values of $\lambda_{max}$ and $P_{max}$ have been derived by fitting the observed polarization in the \textit{B}, \textit{V}, \textit{R}, and \textit{I} passbands to the Serkowski empirical law. The best-fit values of  $P_{max}$  and $\lambda_{max}$ along with the $\sigma_{1}$ for 47 stars, are given in Table \ref{tab:serkowski}. We have not included those stars for which errors in the parameters $P_{max}$ and/or $\lambda_{max}$ were found to be greater than 50\% of the  parameters' values. For the majority of stars,  $\sigma_{1}$ was found to be $<$ 1.6, indicating that the polarization toward the Alessi 1 cluster is mainly due to the foreground interstellar dust grains. However, for three non-member stars (S. No. 6, 24, and 43 in Table \ref{tab:serkowski}) the value of $\sigma_1$ was found to be more than 1.6, which indicates that  these stars feature a non-interstellar component in their measured polarization. Also, for some stars the value of $\lambda_{max}$ was derived to be outside the range of observed wavelengths even for the stars for which $\sigma_{1} < 1.6$. This could be due to the intrinsic polarization of these sources \citep[see also ][]{1975ApJ...196..261S}.  As the exact nature of these stars is not available to us such that we can relate their intrinsic polarization with the properties of the star. Considering member stars of the cluster Alessi 1,  we have derived the mean values of  $\lambda_{max}$  and $P_{max}$ as  $0.59\pm0.04$ $\mu$m and $0.83\pm0.03$\%, respectively. For the polarization, which is dominated by dust particles present in the diffuse ISM, the value of $\lambda_{max}$ lies nearly at 0.55 $\mu$m.

\startlongtable
\begin{deluxetable}{lccccr}
\tablecolumns{6}
\vspace{-0.1in}	
\centering
\tablecaption{Values of $P_{max}$, $\lambda_{max}$, $\sigma_{1}$, membership probability (MP), and color excess $E(B-V)$ for stars in the region of the cluster Alessi 1.	\label{tab:serkowski}}
\tablewidth{20pt}
\setlength\tabcolsep{10.0pt}
\tablehead{
\colhead{S. No.$^\dagger$} & \colhead{$P_{max}$} & \colhead{$\lambda_{max}$} & \colhead{$\sigma_{1}$} & \colhead{MP}& \colhead{$E(B-V)$}\\
\nocolhead{~~} & \colhead{(\%)} & \colhead{($\mu m$)} & \nocolhead{~~} & \nocolhead{~~} &\nocolhead{~~}}
\startdata
1  &   0.89$\pm$0.07  &    0.56$\pm$0.07  &  0.5 & - & - \\ 
2  &   0.90$\pm$0.09  &    0.47$\pm$0.07  &  0.3 & - & - \\		 
3  &   1.46$\pm$0.06  &    0.59$\pm$0.18  &  0.8 & - & - \\			
6  &   0.80$\pm$0.14  &    0.52$\pm$0.22  &  1.9 & - & 0.17 \\ 
7*  &    0.83$\pm$0.02  &    0.49$\pm$0.03  &  0.2 & 0.9 & 0.33 \\ 
9*  &    0.80$\pm$0.10  &    0.66$\pm$0.15  &  0.5 & 0.9 & - \\	
10* &    1.16$\pm$0.08  &    0.72$\pm$0.08  &  0.2 & 1.0 & 0.10 \\
11 &   0.82$\pm$0.09  &    0.53$\pm$0.14  &  0.5 & - & 0.44 \\
12* &    0.78$\pm$0.14  &    0.64$\pm$0.16  &  1.5 & 1.0 & - \\ 
13* &    0.81$\pm$0.05  &    0.60$\pm$0.06  &  0.4 & 0.5 & - \\		
14* &    0.85$\pm$0.08  &    0.52$\pm$0.13  &  0.6 & 0.5 & 0.42 \\  
16 &   0.53$\pm$0.01  &    0.68$\pm$0.01  &  0.1 & - & - \\ 
17 &   0.96$\pm$0.24  &    0.78$\pm$0.19  &  0.7 & - & - \\
18 &   1.13$\pm$0.07  &    0.49$\pm$0.03  &  0.2 & - & - \\ 
22 &   0.81$\pm$0.09  &    0.66$\pm$0.12  &  0.5 & - & - \\		
23 &   1.06$\pm$0.03  &    0.42$\pm$0.03  &  0.3 & - & - \\
24 &   1.98$\pm$0.82  &    1.76$\pm$0.38  &  1.7 & - & 0.21 \\ 		
26 &   0.69$\pm$0.07  &    0.71$\pm$0.12  &  0.3 & - & - \\		
28 &   1.35$\pm$0.06  &    0.67$\pm$0.06  &  0.5 & - & - \\
29 &   0.76$\pm$0.11  &    0.78$\pm$0.15  &  0.3 & - & - \\			
30 &   0.60$\pm$0.07  &    0.59$\pm$0.11  &  0.4 & - & - \\			
34 &   0.81$\pm$0.27  &    0.71$\pm$0.28  &  0.9 & - & 0.13 \\
35 &   0.90$\pm$0.14  &    0.42$\pm$0.09  &  0.5 & - & - \\		
37 &   1.75$\pm$0.23  &    0.26$\pm$0.02  &  0.4 & - & - \\
39 &   0.79$\pm$0.07  &    0.63$\pm$0.10  &  0.5 & - & - \\
41 &   0.72$\pm$0.04  &    0.62$\pm$0.12  &  0.5 & - & - \\
42 &   0.71$\pm$0.06  &    0.50$\pm$0.08  &  0.3 & - & - \\		
43 &   1.10$\pm$0.29  &    0.48$\pm$0.19  &  4.1 & - & - \\
44 &   0.76$\pm$0.16  &    0.46$\pm$0.11  &  0.3 & - & - \\		
45 &   2.19$\pm$0.65  &    0.22$\pm$0.04  &  0.8 & - & - \\
46 &   0.66$\pm$0.01  &    0.45$\pm$0.12  &  1.1 & - & - \\
47* &    0.58$\pm$0.11  &    0.50$\pm$0.20  &  0.6 & 1.0 & 0.26 \\
48* &     0.75$\pm$0.08  &    0.60$\pm$0.11  &  0.5 & 0.9 & - \\
49* &     0.88$\pm$0.07  &    0.55$\pm$0.10  &  0.7 & 1.0 & - \\
50 &   0.98$\pm$0.19  &    0.95$\pm$0.18  &  0.6 & - & - \\		
51* &    0.76$\pm$0.12  &    0.68$\pm$0.23  &  0.6 & 0.8 & - \\
54 &   0.92$\pm$0.06  &    0.70$\pm$0.09  &  0.5 & - & - \\	
55 &   0.94$\pm$0.03  &    0.76$\pm$0.08  &  0.4 & - & - \\
56 &   0.96$\pm$0.01  &    0.47$\pm$0.02  &  0.6 & - & - \\
57* &    0.96$\pm$0.11  &    0.54$\pm$0.15  &  0.6 & - & 0.29 \\	
59 &   1.05$\pm$0.29  &    1.05$\pm$0.23  &  0.7 & - & - \\
60 &   0.89$\pm$0.06  &    0.57$\pm$0.13  &  0.4 & - & - \\		
62 &   1.00$\pm$0.16  &    0.71$\pm$0.25  &  0.8 & - & - \\		
64 &   0.78$\pm$0.03  &    0.40$\pm$0.01  &  0.1 & - & - \\
66 &   0.90$\pm$0.01  &    0.66$\pm$0.01  &  0.1 & - & - \\	
70 &   3.18$\pm$1.32  &    1.44$\pm$0.30  &  0.9 & - & 0.26 \\			
73 &   1.16$\pm$0.03  &    0.79$\pm$0.14  &  0.8 & - & 0.31 \\
\enddata

\vspace{0.1in}	
\tablecomments{$^\dagger$ S. No. in this table are similar to those in the Table \ref{tab:pthe}.\\  * Member stars of the cluster Alessi 1.\\ Membership probabilities are taken from Cantat-Gaudin et al. (2018) and $E(B-V)$ were estimated from \textit{Gaia} $E(G_{BP}-G_{RP})$.}
\end{deluxetable}

The value of $\lambda_{max}$ varies from 0.45 $\mu$m to 0.8 $\mu$m \citep{1975ApJ...196..261S} toward the different lines of sight in the ISM. Furthermore, the mean value of  $\lambda_{max}$ for the cluster Alessi 1 was found to be similar to that found for the location of Alessi 1 in the distribution of $\lambda_{max}$ along with the galactic longitude and latitude in the galactic plane. However, at the galactic latitude near $-13^{o}$, the distribution of $\lambda_{max}$ is not available \citep[see][]{1975ApJ...196..261S}. Therefore, the present observations are useful to fill this gap.

\begin{figure}[h]
	\centering
\includegraphics[width=\columnwidth]{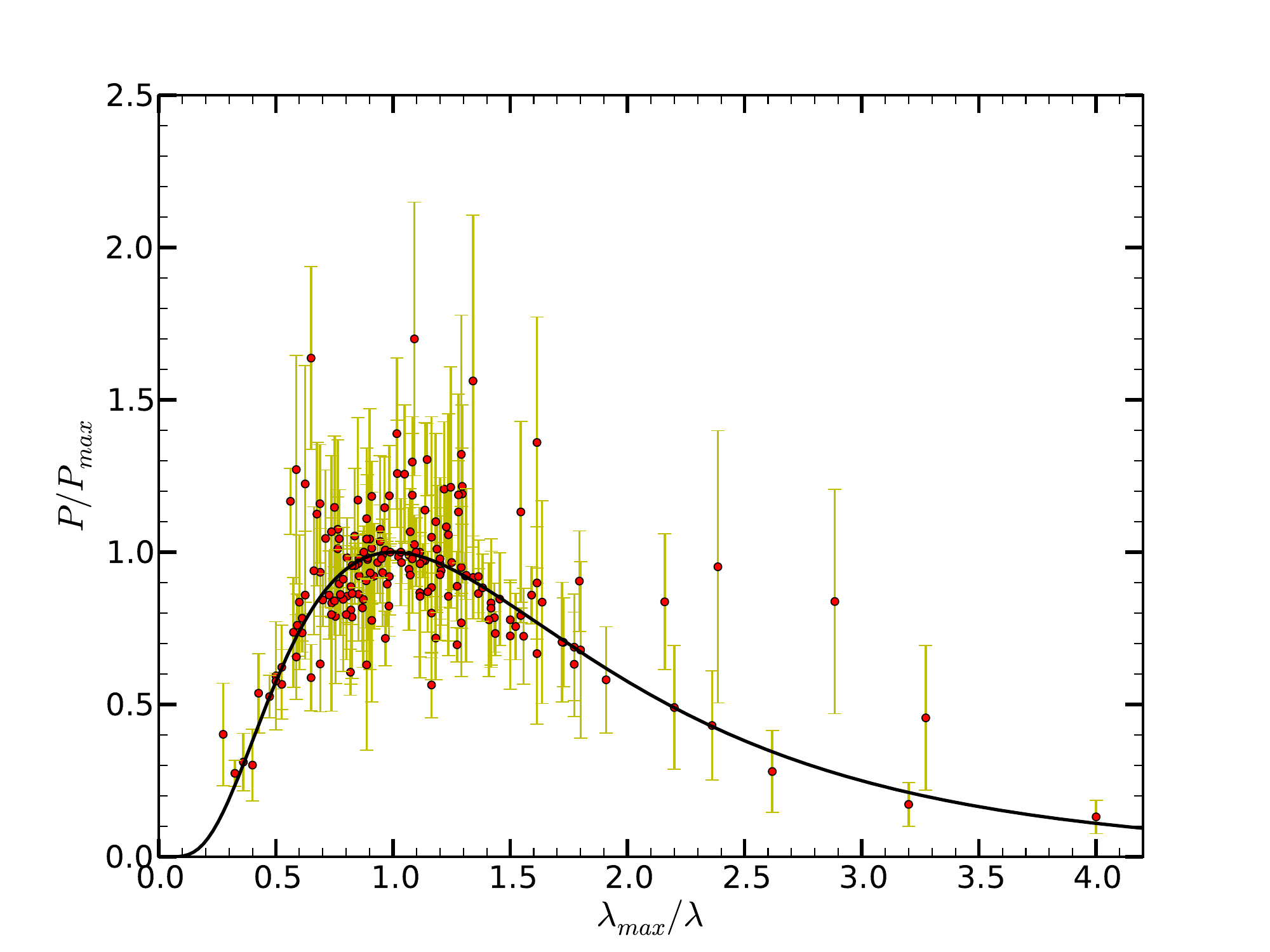}
\caption{Plot of $P/P_{max}$ and $\lambda_{max}/\lambda$. The curve denotes the Serkowski polarization relation for general diffuse ISM.}
  \label{fig:serkowki_plot}
\end{figure}

Figure \ref{fig:serkowki_plot} is a plot of $P/P_{max}$ and $\lambda_{max}/\lambda$. The black curve denotes the Serkowski polarization relation for the general ISM. The Serkowski law for the general ISM was found to be consistent with our observations.  
We have also calculated the value of total to selective extinction toward the  cluster Alessi 1 using the equation $R_{V} = 5.6 \times \lambda_{max}$ \citep{1978A&A....66...57W}. The value of $R_{V}$ toward the line of sight of the cluster  Alessi 1 was found to be $3.3\pm0.2$, which is similar to the normal reddening law, $R_{V} = 3.1$. This indicates that the observed polarizations for the stars in the field of the cluster Alessi 1 are mainly due to the ISM and  the sizes of dust grains are similar to the general ISM grain sizes.

\subsection{Polarization Efficiency}
\label{sec:Polarization Efficiency}
The polarization efficiency is the maximum polarization produced by a given amount of extinction and is measured by the ratio of  $P_{max}$ to the extinction $A_{V}$. It depends on the magnetic field strength, the magnetic field's  orientation, and the degree of alignment of dust grains along the line of sight \citep[see e.g.,][]{2008JQSRT.109.1527V}. 
\begin{figure}[h]
\centering
\subfigure{\includegraphics[width=\columnwidth]{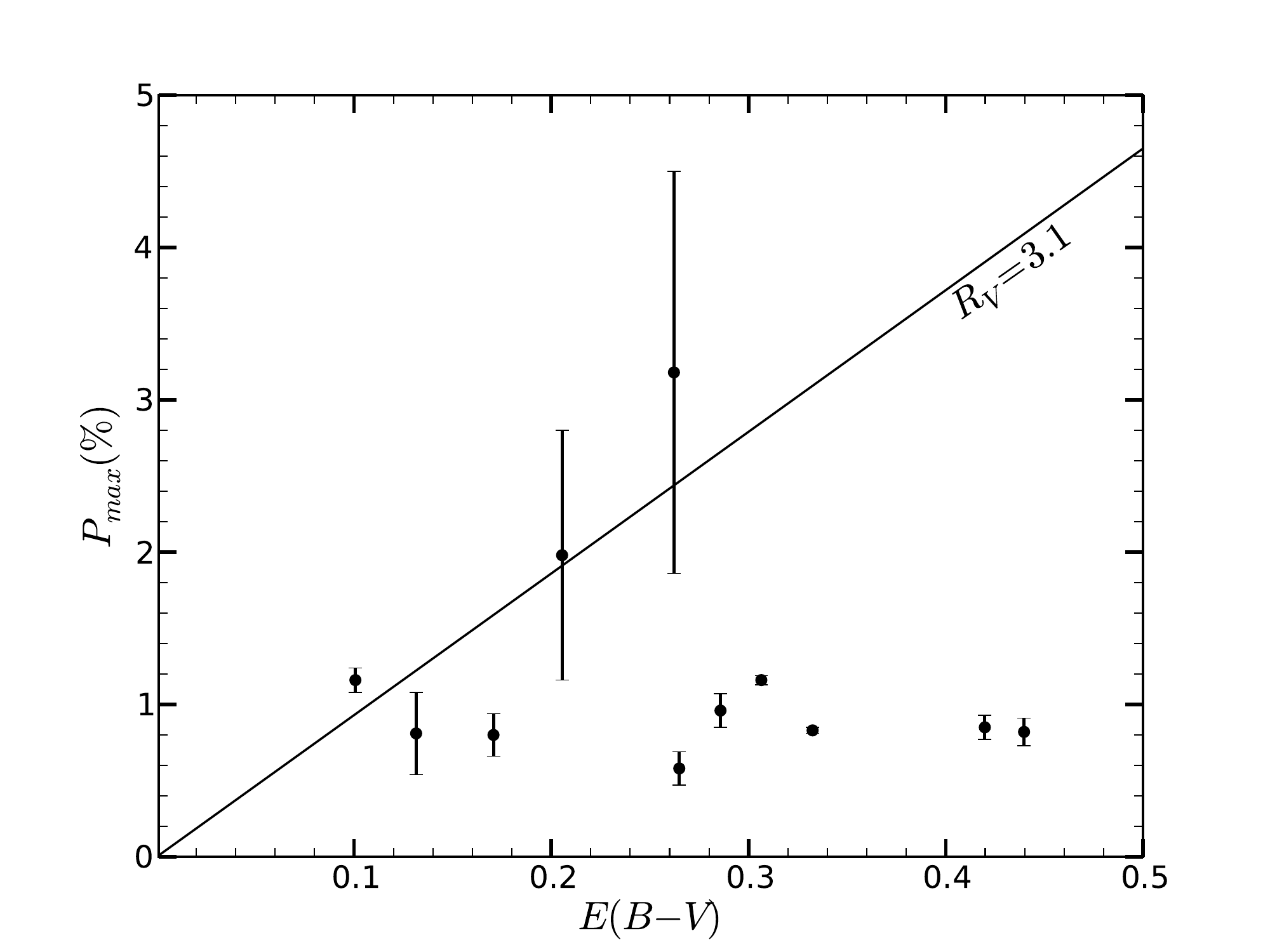}\label{fig:effciencya}}
\subfigure{\includegraphics[width=\columnwidth]{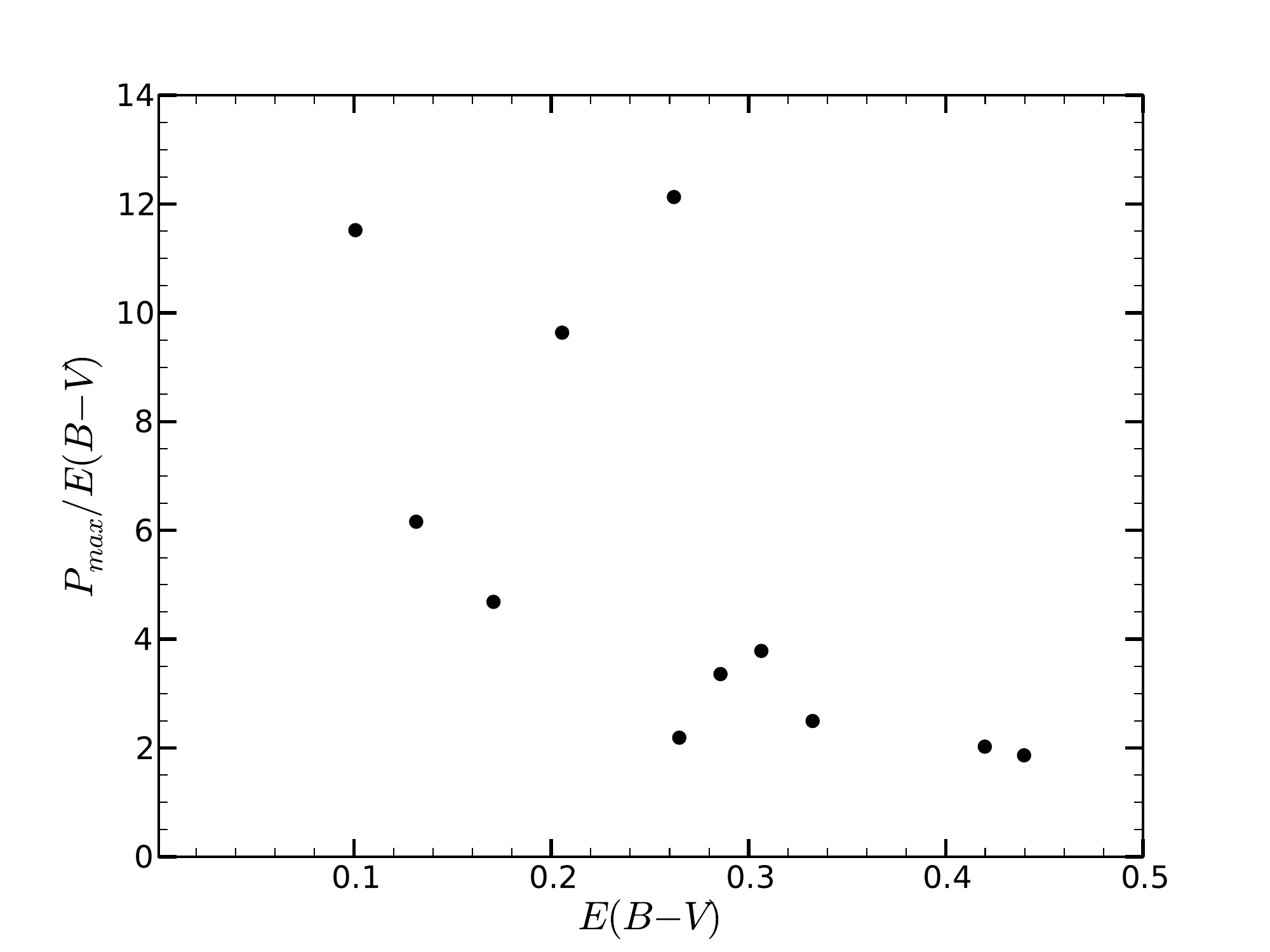}\label{fig:effciencyb}}
\caption{(a) The polarization efficiency curve. The maximum efficiency line using $R_{V} = 3.1$ is drawn. (b) $P_{max}/E(B-V)$ plotted as a function of $E(B-V)$.}
\label{fig:effciency}
\end{figure}

For the diffused ISM, the upper limit on  the polarization efficiency can be estimated by the relation $ P_{max} < 3.0 A_{V} ~\approx ~3 ~R_{V} \times E\left(B - V\right) $ \citep{1956ApJS....2..389H}. Considering the normal reddening law and the average value of $E(B-V)$ for the cluster Alessi 1,  the upper limit of $P_{max}$ was estimated to be 1.5 \%, which is more than the average value of $P_{max}$ derived from the Serkowski law for the member stars. Figure \ref{fig:effciency} shows a polarization efficiency diagram toward the cluster Alessi 1. In Figure \ref{fig:effciencya} we show the variation of $P_{max}$ with $E(B-V)$, where the straight line shows the line of maximum efficiency for $R_{V} = 3.1$. Here, most of the stars are located below the maximum efficiency line. This indicates that net polarization efficiency is due to the ISM. In Figure \ref{fig:effciencyb}, we have plotted $P_{max}/E(B-V)$ as a function of $E(B-V)$. As $E(B-V)$ increases, polarization efficiency was found to decrease. A similar trend was also observed for other clusters where polarization was found to be due to an ISM origin \citep{2008MNRAS.388..105M,2010MNRAS.403.1577M}. This could be due to an increase in the dust grain size or a small change in position angle.

\subsection{Distribution of interstellar matter}
\label{sec:Distribution of interstellar matter}
\begin{figure*}
	\centering
\subfigure[Polarization vectors over-plotted on extinction map]{\includegraphics[width=\columnwidth]{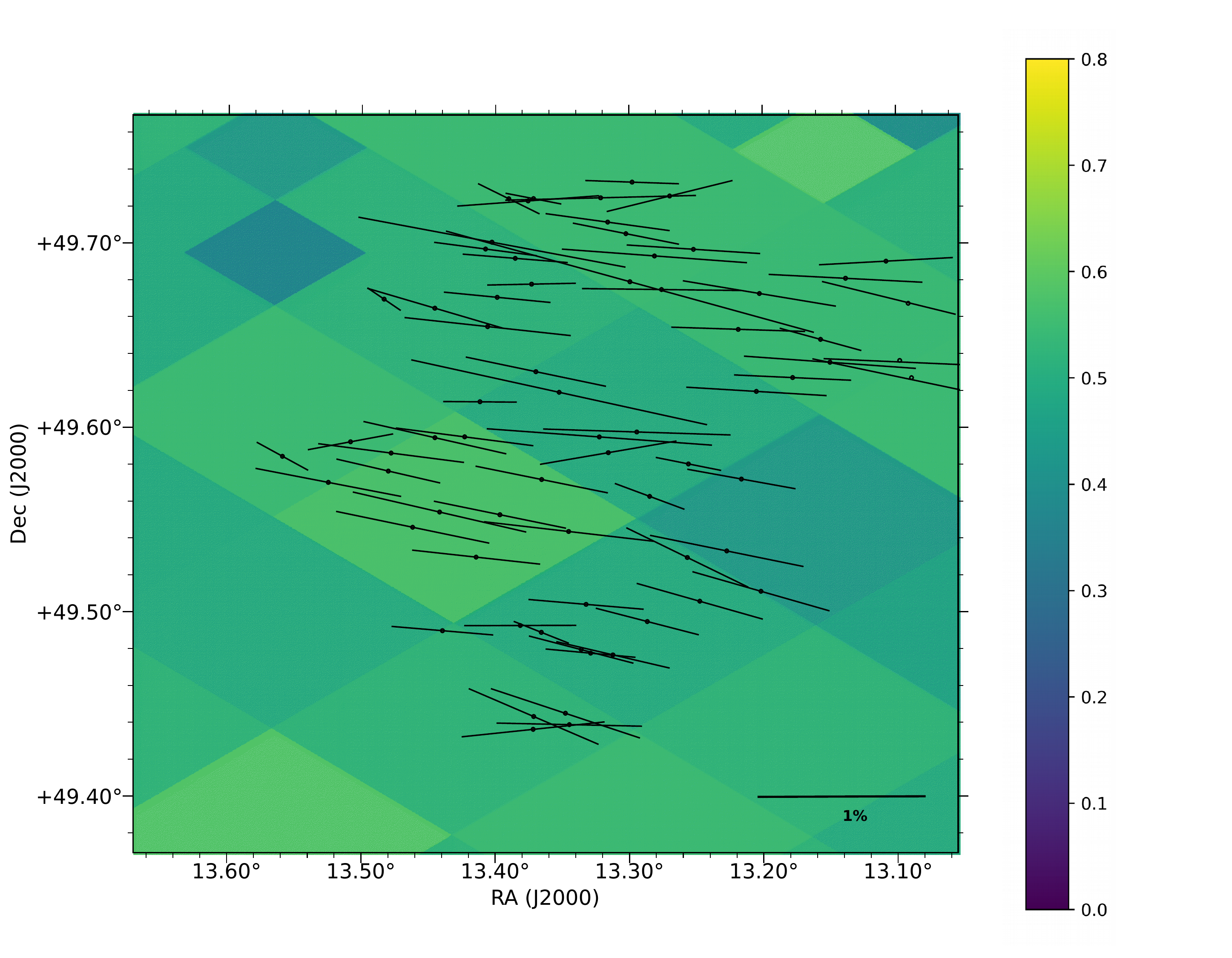}\label{fig:extinction_p}}
\subfigure[Distance versus polarization]{\includegraphics[width=\columnwidth]{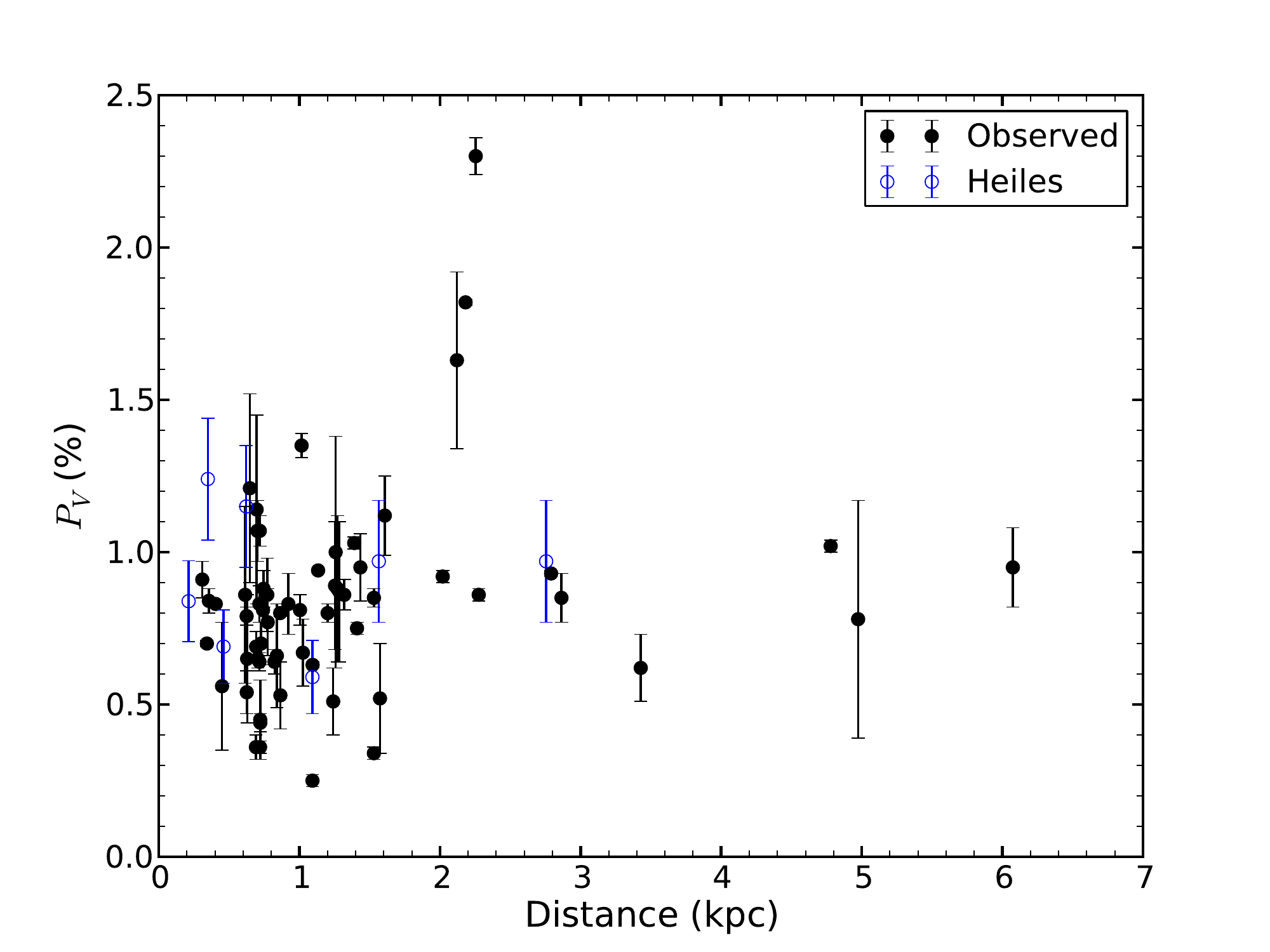}\label{fig:p_distance}}
\caption{(a) Extinction map of the cluster Alessi 1 region  produced by using the Python package dust maps \citep{2018JOSS....3..695G}. The color scale on the right denotes the scale for values of $A_{V}$ for the region. We have overplotted \textit{V}-band polarization results using vectors. The reference for the polarization vector of 1\% polarization is shown at bottom. (b) Variation of degree of polarization in the \textit{V} band with distance.}
    \label{fig:distri_ism_matter}
\end{figure*}

In order to see the distribution of extinction in the cluster Alessi 1, we have used recent extinction maps of \cite{2019ApJ...887...93G} in that region. They have presented a new three-dimensional map of dust reddening based on \textit{Gaia} parallaxes and stellar photometry from the Panoramic Survey Telescope and Rapid Response System and the Two Micron All-Sky Survey covering the sky north of a decl. of $-30^{o}$ (three-quarters of the sky), out to a distance of several kiloparsecs. We have accessed their map using the available Python package dust maps \citep{2018JOSS....3..695G} for the polarimetrically observed region of the cluster Alessi 1.  The polarization vectors in \textit{V} band were overplotted on the extinction map and are shown in  Figure \ref{fig:extinction_p}. The color bar on the right side of the figure shows the range of $A_{V}$ from 0 to 0.8. The reference polarization vector for 1 \% polarization is shown at the bottom of the figure. The extinction toward the region of the cluster Alessi 1 appears to be the same. We do not see any highly extinct region toward this direction. Hence, the origin of polarization must be due to dichroic extinction of starlight by ISM dust grains, as almost all polarization vectors are aligned in the same direction.

It is important to know the distribution of polarization to determine the distribution of interstellar matter toward the particular line of sight. If the radiation from the star encounters to the dust layer at a certain distance then the value of polarization will show a sudden jump at that distance. From this, one can infer the number of  dust layers encountered by radiation of star along the path and also can get an idea about foreground dust concentration toward that line of sight \cite[e.g.,][]{2008MNRAS.388..105M,2010MNRAS.403.1577M,2011MNRAS.411.1418E}. Figure \ref{fig:p_distance} show the variation of degree of polarization in the \textit{V} band with distance. We have also plotted the \textit{V}-band polarization of the stars from the catalog of \citet{2000AJ....119..923H} within a radius of $5^{o}$  around Alessi 1 cluster.  The distances of these stars were taken from  the \textit{Gaia} DR2 catalog. The polarization was found to be almost constant at around 0.8\% with distance. This indicates that no major dust layer is located along the direction of the cluster. The polarization versus distance diagram for many open clusters shows several sudden increments in the value of polarization, indicating the presence of several dust layers in front of them \citep[e.g.,][]{2008MNRAS.388..105M, 2011MNRAS.411.1418E, 2012MNRAS.419.2587E}. Also, using the $Q$-$U$ diagram, several dust layers were found to present in front of the open clusters \citep[e.g.,][]{1999AJ....117.2882W,2004A&A...419..965M,2007A&A...462..621V,2008MNRAS.391..447F,2010MNRAS.403.2041V,2018RMxAA..54..293V}. If we take the case of the cluster Be 59, which is the closest observed cluster for polarization to the Alessi 1, three layers of dust are found to present in front of it \citep{2012MNRAS.419.2587E}. 

\section{Summary}\label{sec:sum}
We have carried out a polarimetric study of  73 stars toward the cluster Alessi 1. The age, distance, and $E(B-V)$ of the cluster Alessi 1 are estimated to be $0.8\pm0.1$ Gyr, $673\pm98$ pc, and $0.16\pm0.03$ mag, respectively, using the archival data from \textit{Gaia}. A total of 66 stars are found to be members of the cluster Alessi 1 using an astrometric approach, out of which 15 member stars are observed polarimetrically. The average value of $P$ and $\theta$ for member stars of the cluster Alessi 1 are found to be $0.77\pm0.01$\% and $84^{o}.4\pm0^{o}.1$, $0.70\pm0.01$\% and $75^{o}.2\pm0^{o}.2$, $0.76\pm0.01$\% and $74^{o}.9\pm0^{o}.1$, and $0.96\pm0.02$\% and  $83^{o}.7\pm0^{o}.6$ in the \textit{B}, \textit{V}, \textit{R}, and \textit{I} bands, respectively.  The distribution of the dust grain toward the direction of Alessi 1 is found to be similar as general ISM as $\lambda_{max}$ is found to be $0.59\pm0.04$ $\mu$m. The average value of maximum polarization is found to be  $0.83\pm0.03$ \% when considering only the member stars.  It is found that polarization toward the Alessi 1 cluster is dominated by foreground dust grains, which are probably aligned by the Galactic magnetic field.

\section*{Acknowledgment}
We thank the referee for reading our paper and the useful comments.
Part of this work has made use of data from the European Space Agency (ESA) mission
{\it Gaia} (\url{https://www.cosmos.esa.int/gaia}), processed by the {\it Gaia}
Data Processing and Analysis Consortium (DPAC,
\url{https://www.cosmos.esa.int/web/gaia/dpac/consortium}). Funding for the DPAC
has been provided by national institutions, in particular, the institutions
participating in the {\it Gaia} Multilateral Agreement. 
This research made use of the cross-match service provided by CDS, Strasbourg.

\bibliography{ref}{}
\bibliographystyle{aasjournal}

\end{document}